\documentclass[10pt,journal,compsoc]{IEEEtran}

\ifCLASSOPTIONcompsoc
  \usepackage{cite}
\else
  \usepackage{cite}
\fi
\ifCLASSINFOpdf
\else
\fi
\usepackage{multirow}

\usepackage{subfigure}

\usepackage{comment}
\usepackage{amsmath,amssymb,amsfonts}
\usepackage{algorithmic}
\usepackage{graphicx}
\usepackage{textcomp}
\usepackage{xcolor}
\usepackage{mathtools}
\usepackage{float}
\usepackage[ruled,vlined]{algorithm2e}
\usepackage{bbding} %导入包
\usepackage{caption}  
\usepackage{bm}
\usepackage{xspace}

\renewcommand{\vec}[1]{\mathbf{#1}}

\begin{document}
%
% paper title
% Titles are generally capitalized except for words such as a, an, and, as,
% at, but, by, for, in, nor, of, on, or, the, to and up, which are usually
% not capitalized unless they are the first or last word of the title.
% Linebreaks \\ can be used within to get better formatting as desired.
% Do not put math or special symbols in the title.
\title{ Multi-Objective Recommendation in the Era of Generative AI: A Survey of  Recent Progress and Future Prospects}

\author{Zihan Hong$^*$,
        Yushi Wu$^*$,
        Zhiting Zhao,
        Shanshan Feng$^\dagger$,
        Jianghong Ma$^\dagger$,
        Jiao Liu,
        Tianjun Wei
        %Yew Soon Ong

        % and~Yew-Soon Ong,~\IEEEmembership{Fellow,~IEEE}
        % <-this % stops a space
\IEEEcompsocitemizethanks{

% \IEEEcompsocthanksitem \#These authors contributed equally.

% \IEEEcompsocthanksitem *These authors are corresponding authors.

\IEEEcompsocthanksitem Zihan Hong is an undergraduate student from Harbin Institute of Technology, Shenzhen, China. Email: zihanhong82@gmail.com

\IEEEcompsocthanksitem Yushi Wu is an undergraduate student from Harbin Institute of Technology, Shenzhen, China. Email: wuyushi2004@gmail.com

\IEEEcompsocthanksitem Zhiting Zhao is a graduate student from Harbin Institute of Technology, Shenzhen, China. Email: 24S151185@stu.hit.edu.cn

\IEEEcompsocthanksitem Shanshan Feng is with Wuhan University, Wuhan, China.
E-mail: victor\_fengss@whu.edu.cn

\IEEEcompsocthanksitem Jianghong Ma is with Harbin Institute of Technology, Shenzhen, China. Email: majianghong@hit.edu.cn%

\IEEEcompsocthanksitem Jiao Liu is with Nanyang Technological University, Singapore. Email: jiao.liu@ntu.edu.sg%

\IEEEcompsocthanksitem Tianjun Wei is with Nanyang Technological University, Singapore. Email: tjwei2-c@my.cityu.edu.hk%

%\IEEEcompsocthanksitem Yew-Soon Ong is with the Agency for Science, Technology and Research (A*STAR), Singapore and also with Nanyang Technological University, Singapore. E-mail: asysong@ntu.edu.sg

% \IEEEcompsocthanksitem \#contributed equally, *Corresponding authors.

}% <-this % stops an unwanted space

\thanks{Manuscript received April 19, 2005; revised August 26, 2015.
\\
($^*$Both authors contributed equally to this research, $^\dagger$Corresponding author)
}}

% note the % following the last \IEEEmembership and also \thanks - 
% these prevent an unwanted space from occurring between the last author name
% and the end of the author line. i.e., if you had this:
% 
% \author{....lastname \thanks{...} \thanks{...} }
%                     ^------------^------------^----Do not want these spaces!
%
% a space would be appended to the last name and could cause every name on that
% line to be shifted left slightly. This is one of those "LaTeX things". For
% instance, "\textbf{A} \textbf{B}" will typeset as "A B" not "AB". To get
% "AB" then you have to do: "\textbf{A}\textbf{B}"
% \thanks is no different in this regard, so shield the last } of each \thanks
% that ends a line with a % and do not let a space in before the next \thanks.
% Spaces after \IEEEmembership other than the last one are OK (and needed) as
% you are supposed to have spaces between the names. For what it is worth,
% this is a minor point as most people would not even notice if the said evil
% space somehow managed to creep in.

% The paper headers
\markboth{Journal of \LaTeX\ Class Files,~Vol.~14, No.~8, August~2015}%
{Shell \MakeLowercase{\textit{et al.}}: Bare Demo of IEEEtran.cls for Computer Society Journals}

% for Computer Society papers, we must declare the abstract and index terms
% PRIOR to the title within the \IEEEtitleabstractindextext IEEEtran
% command as these need to go into the title area created by \maketitle.
% As a general rule, do not put math, special symbols or citations
% in the abstract or keywords.
\IEEEtitleabstractindextext{
\begin{abstract}
With the recent progress in generative artificial intelligence (Generative AI), particularly in the development of large language models, recommendation systems are evolving to become more versatile. Unlike traditional techniques, generative AI not only learns patterns and representations from complex data but also enables content generation, data synthesis, and personalized experiences. This generative capability plays a crucial role in the field of recommendation systems, helping to address the issue of data sparsity and improving the overall performance of recommendation systems. Numerous studies on generative AI have already emerged in the field of recommendation systems. Meanwhile, the current requirements for recommendation systems have surpassed the single utility of accuracy, leading to a proliferation of multi-objective research that considers various goals in recommendation systems. However, to the best of our knowledge, there remains a lack of comprehensive studies on multi-objective recommendation systems based on generative AI technologies, leaving a significant gap in the literature. Therefore, we investigate the existing research on multi-objective recommendation systems involving generative AI to bridge this gap. We compile current research on multi-objective recommendation systems based on generative techniques, categorizing them by objectives. Additionally, we summarize relevant evaluation metrics and commonly used datasets, concluding with an analysis of the challenges and future directions in this domain. 
\end{abstract}

% Note that keywords are not normally used for peerreview papers.
}

% make the title area
\maketitle

% To allow for easy dual compilation without having to reenter the
% abstract/keywords data, the \IEEEtitleabstractindextext text will
% not be used in maketitle, but will appear (i.e., to be "transported")
% here as \IEEEdisplaynontitleabstractindextext when the compsoc 
% or transmag modes are not selected <OR> if conference mode is selected 
% - because all conference papers position the abstract like regular
% papers do.
\IEEEdisplaynontitleabstractindextext
% \IEEEdisplaynontitleabstractindextext has no effect when using
% compsoc or transmag under a non-conference mode.

% For peer review papers, you can put extra information on the cover
% page as needed:
% \ifCLASSOPTIONpeerreview
% \begin{center} \bfseries EDICS Category: 3-BBND \end{center}
% \fi
%
% For peerreview papers, this IEEEtran command inserts a page break and
% creates the second title. It will be ignored for other modes.
\IEEEpeerreviewmaketitle

\IEEEraisesectionheading{\section{Introduction}\label{sec:introduction}}
\IEEEPARstart
In the era of big data, recommendation systems have emerged as indispensable tools for addressing information overload, enabling users to discover valuable content efficiently. Widely applied across domains such as music, news, and job recommendations \cite{dong2024musechat, li2024review, patil2023survey}, these systems enhance user experiences by filtering vast information streams. Their evolution spans decades, from early collaborative filtering \cite{schafer2007collaborative, herlocker2004evaluating, papadakis2022collaborative, koren2021advances} and content-based methods \cite{lops2011content, javed2021review} to hybrid models \cite{dong2017hybrid}, graph neural networks-based \cite{fan2020graph} and deep learning-based \cite{zhang2019deep, naumov2019deep}, continuously advancing to meet growing demands for personalization and scalability.

Recent advances in generative AI have significantly transformed the landscape of recommendation systems.
As noted by \cite{wang2023generative}, generative recommendation systems based on generative technologies represent an emerging research direction in this field. Techniques such as Generative Adversarial Networks (GANs) \cite{goodfellow2014generative}, Variational Autoencoders (VAEs) \cite{liang2024survey}, diffusion models \cite{wang2023diffusion}, and large language models (LLMs) \cite{wu2024survey} enable richer data synthesis and deeper context understanding. Among these, LLMs are particularly effective at handling multimodal data (text, images, videos) and generating context-aware recommendations, offering unprecedented flexibility. Unlike traditional models that predict user preferences based on historical data, generative models can simulate user interactions, augment sparse datasets, and create personalized content. These capabilities open new avenues for innovation in recommendation paradigms.

Generative models show promise in the field of recommendation system. Current researches primarily focus on single-objective tasks, such as improving accuracy through synthetic data generation or leveraging LLMs for explainability. However, this narrow focus on accuracy risks creating a filter bubble \cite{nguyen2014exploring}, where users are confined to repetitive or homogeneous content, stifling exploration and diminishing long-term engagement. Given the advanced reasoning and comprehension capabilities of generative AI, its applications in multi-objective recommendation are also promising. 
The research community has extensively explored multi-objective recommendation systems that balance competing goals within traditional recommender system frameworks \cite{jannach2022multi, Neurocomputing_zheng2022survey, FBD_jannach2023survey, kaminskas2016diversity}. However, the development of multi-objective recommendation systems leveraging emerging generative AI techniques remains underexplored. 
Therefore, the integration of multi-objective optimization into generative recommendation systems warrants further comprehensive investigation.

To address this gap, we systematically investigate existing studies that employ generative technologies to enable multi-objective recommendation systems.
It is crucial to emphasize that, we consider that any discussion of additional objectives in recommendation systems, such as diversity, serendipity, or fairness, implicitly assumes accuracy as a foundational requirement. Thus, we define \emph{multi-objective recommendation systems (MORS)} as those achieving optimization of beyond-accuracy metrics.

Our survey identifies \emph{diversity}, \emph{serendipity}, \emph{fairness}, and \emph{safety} as the primary beyond-accuracy objectives in generative recommendation systems, along with additional objectives such as \emph{novelty}, \emph{controllability}, \emph{efficiency}, and \emph{robustness}. Meanwhile, we focus on four generative techniques widely adopted in recommendation systems, which are \emph{GANs}, \emph{diffusion models}, \emph{VAEs}, and \emph{LLMs}. For each objective, we critically review the underlying model architectures and evaluation metrics prevalent in generative recommendation systems, and also summarize relevant development challenges, aiming to provide foundational insights for future research in multi-objective generative recommendation.

The main contributions are summarized as follows:

\begin{itemize}
    \item This paper is the first comprehensive survey that integrates generative AI (GANs, VAEs, diffusion models and LLMs) with MORS. It proposes a object-oriented classification framework, systematically reviewing advancements and limitations in model architecture, optimization strategies, and evaluation metrics across four key objectives: diversity, serendipity, fairness, and security.
    \item We systematically summarize specialized evaluation metrics and corresponding benchmark datasets for different objective domains such as fairness and serendipity, providing standardized references for experimental design.
    \item We also discuss key challenges in generative MORS and outline future research directions. We highlight the need for improved evaluation metrics, advanced strategies for LLMs, and the integration of multiple generative techniques to enhance recommendation quality. Additionally, we emphasize the importance of interdisciplinary collaboration, drawing from ethics and sociology to develop fairer and more transparent systems. These insights provide a foundation for further exploration and innovation in both academia and industry. 
\end{itemize}

\noindent \textbf{Paper Overview.} \quad In \textbf{Section 2}, we primarily summarize the existing literature on recommendation systems, generative recommendation systems, and MORS, establishing the background for our study. In \textbf{Section 3}, we introduce the four major generative techniques covered in our survey. After that, \textbf{Section 4}, the core of the paper, surveys MORS based on generative techniques. We categorize them according to the beyond-accuracy objective and examine the related definitions, models, and evaluation metrics. In the following \textbf{Section 5}, we summarize the common datasets used for each objective in the context of generative recommendation. In \textbf{Section 6}, we summarize the current main challenges for each key super-accuracy metric. Finally, in \textbf{Section 7}, we provide a summary of the entire paper.

\graphicspath{ {image/} }  %image为文件夹名，可以在左侧自己创建文件夹

\section{Related Work}
We review existing literature on conventional and generative recommendation systems, as well as MORS, to contextualize and motivate the present study.

\subsection{Traditional Recommendation System}
Recommendation systems emerged in response to the information overload problem, where users are confronted with an overwhelming amount of information and struggle to discover items or content, such as products, movies, music, or articles, that align with their interests. With the advent of big data, recommendation systems have become indispensable tools for filtering and surfacing relevant content to users.

Currently, algorithms in the field of recommendation systems are generally divided into three main categories: content-based filtering, collaborative filtering, and hybrid filtering methods \cite{roy2022systematic, isinkaye2015recommendation, kumar2018recommendation}. \emph{Content-based filtering algorithms} recommend items by analyzing item features and matching them with user profiles. \emph{Collaborative filtering algorithms}, on the other hand, leverage similarity measurements between users (or items) to identify those with similar preferences to the target user (or item). \emph{Hybrid filtering algorithms} integrate two or more recommendation techniques to overcome the limitations of individual approaches.  

Regardless of the algorithm used, a crucial step is the deep exploration of the input data. With advances in AI such as deep learning \cite{zhang2019deep, naumov2019deep} and graph neural networks \cite{wu2022graph, gao2023survey}, increasingly sophisticated models have been developed to better extract data features, capture complex user–item interactions, and integrate heterogeneous data sources. These innovations have significantly improved the modeling of user preferences and item characteristics, helping to address common challenges in recommendation systems, such as the cold-start problem and data sparsity.

We could also note that numerous surveys have already been conducted on the application of AI technologies in recommendation systems. For instance, the article \cite{zhang2021artificial} systematically surveys eight key areas of AI technologies and their applications in recommendation systems. It provides a comprehensive overview of state-of-the-art AI algorithms, covering models, methods, and applications. Specifically, it delves into deep neural networks, transfer learning, active learning, reinforcement learning, fuzzy techniques, evolutionary algorithms, natural language processing, and computer vision. Similarly, Masciari \textit{et al.} \cite{masciari2024systematic} discusses commonly used AI techniques in recommendation systems, including convolutional neural networks (CNNs), collaborative filtering, long- and short-term memory (LSTM), decision trees, and Naïve Bayes. These traditional AI-based recommendation systems primarily utilize AI technologies for mining existing data, including feature extraction and enriching data types, thereby significantly enhancing the performance of recommendation systems.

Now, with the rapid evolution of AI technologies, generative techniques emerge as a transformative force in recommendation systems, demonstrating superior performance in data mining and representation learning compared to traditional AI methods. Additionally, they also excel in generating creative content. Consequently, exploring generative recommendation systems has emerged as a pivotal research frontier in both academia and industry.

\subsection{Generative Recommendation System}

In recent years, generative techniques have been widely applied in recommendation systems, attracting increasing attention \cite{wu2024survey, bharadhwaj2018recgan, liu2020deep, zheng2023generative, gao2024generative, li2023gpt4rec, walker2022recommendation}. Therefore, there have been some studies which have investigated the development of generative techniques in recommendation systems \cite{wang2023generative, liang2024survey, wu2024survey, gao2021recommender, li2023large, zhao2023recommender, deldjoo2024review, li2024survey, ayemowa2024analysis, wang2024enhanced, lin2024survey, vats2024exploring, lin2025can}. We must first emphasize that in this article, the term "Generative Recommendation" or "Generative Recommendation System" specifically refers to recommendation systems employing generative technologies. This definition differs from studies that narrowly characterize generative recommendation as directly producing recommendation content in a single step without separately calculating ranking scores for individual candidates. Our operational definition thus encompasses a broader conceptual scope.

Through a systematic review of relevant literature, we have identified that surveys on generative recommendation commonly focus on four generative techniques, including GANs, VAEs, Diffusion Models and LLMs. A detailed discussion of these technologies in recommendation systems will be presented in Section 3. 

For \emph{LLMs}, Li \textit{et al.} \cite{li2023large} primarily focus on directly employing LLMs for recommendation generation, deliberately excluding LLM-based discriminative recommendation approaches. \cite{xu2024large} focuses on the impact of LLMs in the domain of information extraction. Zhao \textit{et al.} \cite{zhao2023recommender} provide a comprehensive review of LLM-based recommender systems, discussing pre-training, fine-tuning, and prompting paradigms. \cite{li2024survey} systematically reviews the latest developments in generative search and recommendation and proposes a unified generative framework. Paper \cite{wu2024survey} categorizes existing LLM-based recommendation systems into two paradigms: Discriminative LLM for Recommendation (DLLM4Rec) and Generative LLM for Recommendation (GLLM4Rec), and is the first to systematically classify the latter. Paper \cite{lin2025can} summarizes existing research studies from two orthogonal aspects: where and how to adapt LLMs to recommendation systems. Paper \cite{liang2024survey} summarizes the applications of LLMs in recommendation systems across various domains, and also delves into the application of LLMs with prompt engineering and fine-tuning. For \emph{GANs}, Gao et al. \cite{gao2021recommender} review the application of GANs in recommendation systems, including solving the data noise and data sparsity issue. For \emph{diffusion models}, paper \cite{lin2024survey} conducts a survey on the application of diffusion models in recommendation systems, categorizing existing work based on different use cases. For \emph{VAEs}, Liang et al. \cite{liang2024survey} provide a dedicated summary of recent VAE-based recommendation algorithms.

Additionally, Ayemowa \textit{et al.} \cite{ayemowa2024analysis} focus on two generative techniques which are GANs and VAEs and review the selected studies techniques, models, datasets and metrics. In \cite{deldjoo2024review}, Deldjoo \textit{et al.} survey a broader spectrum of the generative models used in recommendation systems, including interaction-driven generative models, LLM-based natural language recommenders, and multimodal recommendation models.

While research on generative techniques in recommendation systems is extensive, the majority of existing studies  primarily emphasizing improvements in recommendation accuracy, neglecting other crucial objectives in recommender systems. Overall, multi-objective recommendation, which considers goals such as diversity, fairness, or serendipity, remains underexplored in the context of generative models.

\subsection{Multi-Objective Recommendations}
Multi-objective recommendation systems represent an evolution of traditional recommender systems by addressing the limitations of single-objective models by optimizing multiple, often conflicting, objectives simultaneously. These systems aim to provide a more nuanced and comprehensive recommendation by balancing different quality metrics such as accuracy, diversity, serendipity, safety and others. 

Numerous studies have contributed to defining and formalizing these objectives in the context of recommendation systems. Alhijawi \textit{et al.} \cite{alhijawi2022survey} identify five main objectives of recommendation systems, which are relevance, diversity, novelty, coverage, and serendipity. They also provide a detailed overview of various evaluation metrics for recommendation systems, including their definitions and mathematical formulations. Meanwhile, paper \cite{jannach2022multi} proposes a classification framework for multi-objective recommendation settings. It categorizes objectives into five groups: recommendation quality objectives, multi-stakeholder objectives, temporal objectives, user experience objectives, and system objectives. In addition to classifying and summarizing all objectives of MORS, there are also some studies that focus on specific objectives within multi-objective recommendation to analyze and refine their definitions. For example, paper \cite{kunaver2017diversity} explores diversity, paper \cite{mendoza2020evaluating} examines novelty, paper \cite{kotkov2016survey, kotkov2023rethinking} discuss serendipity, paper \cite{himeur2022latest} addresses security and privacy concerns, and paper \cite{wang2023survey, chen2023bias} provide a systematic review of fairness and bias issues. 

Beyond the definition of objectives, multi-objective recommendation techniques are also a common focus of research in this field. In traditional multi-objective recommendation, the common used techniques are multi-objective optimization techniques \cite{zaizi2023comparative, perera2023multiple, zaizi2024towards, zhang2022community, zhang2021balancing, zhou2023dynamic, zhou2024diversified,zaizi2023multi}. By employing these techniques, MORS can generate a set of recommendations that offer a trade-off among various objectives, enhancing personalization and user satisfaction. These algorithms can be divided into three categories: (1) Deterministic approaches; (2) Stochastic-based techniques; (3) Stochastic learning approaches \cite{zaizi2023multi}. However, \cite{zaizi2023multi} points out that the three categories of methods face several challenges. %including: 1) how to select the appropriate multi-objective optimization technique based on the given problem scenario; 2) how to effectively obtain real-world datasets that accurately reflect user preferences and behaviors; 3)
% how to better evaluate the trade-offs between multiple objectives, ensuring that the model optimally balances competing goals; 4) how to appropriately balance the trade-offs between different objectives, taking into account their relative importance in the context of the specific recommendation system. At the same time, 
The drawbacks of different multi-objective techniques have been thoroughly summarized in the \cite{zaizi2023multi}, and are therefore not repeated here. This highlights a growing need for new technological methods in this field. 

Subsequently, a potential development direction is explored in \cite{liu2024large}. This paper proposes a new framework that combines large language models (LLMs) with traditional multi-objective evolutionary algorithms to enhance the algorithm's search and generalization capabilities. This work demonstrates the potential of LLMs, as one of the generative technologies, in advancing the field of MORS. Motivated by this, we aim to systematically investigate the role of generative techniques in the development of MORS.

\begin{table*}[h]
    \centering
    \renewcommand{\arraystretch}{1.5} % 调整行高
    \setlength{\tabcolsep}{15pt} % 调整列间距
    % \caption{Overview of related surveys and our survey}
    \caption{Comparison between this survey and existing surveys}
    \begin{tabular}{cccccc}
    \hline  \textbf{References} & \textbf{LLMs} & \textbf{GANs} & \textbf{VAEs} & \textbf{Diffusion models} & \textbf{Multi-objecctive} \\ \hline\cite{li2023large,xu2024large,xu2024large,zhao2023recommender,li2024survey,wu2024survey,lin2025can,vats2024exploring} & \Checkmark & \XSolidBrush &\XSolidBrush & \XSolidBrush & \XSolidBrush \\
     \cite{gao2021recommender} & \XSolidBrush & \Checkmark & \XSolidBrush &\XSolidBrush & \XSolidBrush\\
     \cite{liang2024survey}  & \XSolidBrush  & \XSolidBrush & \Checkmark &\XSolidBrush & \XSolidBrush\\
     \cite{ayemowa2024analysis}  & \XSolidBrush  & \Checkmark  & \Checkmark  &\XSolidBrush & \XSolidBrush\\
     \cite{deldjoo2024review}   & \Checkmark  & \Checkmark  & \Checkmark  &\Checkmark & \XSolidBrush\\
     \cite{zaizi2023comparative, perera2023multiple, zaizi2024towards, zhang2022community, zhang2021balancing, zhou2023dynamic, zhou2024diversified,zaizi2023multi}  & \XSolidBrush &\XSolidBrush & \XSolidBrush & \XSolidBrush & \Checkmark \\ \hline
     Our Survey & \Checkmark  & \Checkmark  & \Checkmark  &\Checkmark & \Checkmark\\ \hline 
    \end{tabular}
    \label{tab:comparison}
    % \vspace{6pt}
\end{table*}

\subsection{Differences from Existing Surveys}

As discussed above, the existing research articles on traditional recommendation systems, generative recommendation systems, and MORS are already extensive. However, there are still some limitations. 
First, existing surveys on generative recommendation systems have not explored their development in the context of MORS. Second, surveys focused on MORS primarily address traditional recommendation approaches and largely overlook recent developments in generative techniques.
%(1) Current surveys on generative recommendation systems have not explored their development in the context of MORS; (2) Surveys related to MORS primarily focus on traditional multi-objective recommendation algorithms. 

%Given the current potential of generative technologies in MORS, our survey aims to survey MORS with generative technologies, in order to fill this research gap. 
Given the increasing potential of generative AI in enhancing MORS, this survey aims to provide a comprehensive review of generative multi-objective recommendation systems. Specifically, we focus on how generative AI techniques can be leveraged to achieve beyond-accuracy objectives.
Overall, research at the intersection of generative AI and MORS remains in its early stages, with relatively less studies available compared to those focusing on single-objective settings.
 Considering the limited research in this area, our survey adopts an objective-centric organization, categorizing existing studies based on the specific beyond-accuracy objectives they address. This organization is elaborated in Section 4. 
 
 % To highlight the differences between our research and existing survey studies, we compare relevant works from the perspectives of multi-objective optimization and generative AI techniques , with the results summarized in TABLE 1.

 To clarify the scope and novelty of this survey, we also compare representative related surveys across two dimensions: generative AI techniques and multi-objective. This comparison, summarized in Table \ref{tab:comparison}, highlights the distinctiveness of our survey in addressing both fronts.

%Additionally, we will primarily focus on four major generative technologies, including Generative Adversarial Networks (GANs), Diffusion Models, Variational Autoencoders (VAEs), and Large Language Models (LLMs), which we will elaborate on in detail in the \textbf{TECHNOLOGY} section. 
%\input{Sections/4.Objectives}
\section{Generative Technology for Recommendation}
This section introduces four generative AI techniques frequently employed in recommendation systems, as previously outlined in our discussion.

\subsection{Generative Adversarial Networks}

GANs are generative models designed to learn data distributions through adversarial training and generate new samples. It consists of two primary components \cite{goodfellow2014generative}: (1) \emph{Generator}, which maps random noise to synthetic samples that resemble real data, with the objective of deceiving the discriminator; (2) \emph{Discriminator}, which distinguishes between real samples and those generated by the generator, aiming to improve classification accuracy.

By leveraging adversarial dynamics between the generator and discriminator, GANs enable implicit learning of complex data distributions and the generation of high-quality samples. They have been widely applied in image synthesis \cite{brock2018large, zhan2019spatial, qin2020gan}, style transfer \cite{azadi2018multi, xu2021drb}, data augmentation \cite{tran2021data, frid2018synthetic, mariani2018bagan}, missing data imputation \cite{lee2019collagan, yoon2018gain}, and few-shot learning \cite{zhang2018metagan}. 
% However, GANs often suffer from issues such as mode collapse, where the generator produces limited sample diversity, and difficulties in achieving convergence. What's more,their training involves non-convex optimization and game-theoretic dynamics, requiring empirical tuning and heuristics to ensure stable learning.
%The generative capability of GANs is often used to learn data distributions and generate new data. 
In recommendation systems, this technique has been commonly applied to address issues related to data sparsity and cold start problems \cite{wang2021siamese, woo2022conditional, woo2022conditional, chen2024collaborative, shafqat2022hybrid}.
%For example, the paper\cite{wang2021siamese} to learn the data distribution characteristics of high-dimensional sparse matrices.% By doing so, it estimates unknown entries, addressing the challenge of estimating missing values in high-dimensional sparse matrices within recommendation systems.
%Another paper \cite{woo2022conditional} employs two conditional GANs to augment data for cold-start users. %This approach enhances collaborative filtering recommendation algorithms, ultimately improving recommendation accuracy.
%Similarly, the paper \cite{chen2024collaborative} also proposes a collaborative filtering algorithm based on GANs and attention mechanisms to address data sparsity and cold start problems. Additionally, the paper \cite{shafqat2022hybrid} introduces a hybrid GAN-based method that generates synthetic data to oversample minority-class data. %This approach addresses data imbalance in recommendation systems, leading to significant improvements in recommendation accuracy. 
Furthermore, GAN technology can be combined with other generative techniques, such as VAEs, to enhance recommendation performance. For instance, in paper \cite{zhou2023vcgan}, the generator of GAN is implemented using a VAE. The rapid development of GANs also has led to the emergence of many variants, including Conditional GANs (cGANs), CNNs, Wasserstein GANs (WGANs), FairGNN, Social Trust GANs, and the MRNGAN model. 
These models are comprehensively summarized in the article \cite{ayemowa2024analysis}.

Overall, GAN-based approaches enhance recommendation performance by generating high-quality synthetic data, improving user feature modeling, optimizing item distribution, and integrating with other advanced techniques.

\subsection{Diffusion Models}
Diffusion models typically consist of two main processes \cite{ho2020denoising}: a forward process and a reverse process. The \emph{forward process} involves gradually introducing noise from the initial state to simulate the diffusion of information, while the \emph{reverse process} removes noise step by step to recover the original information.

In the context of recommendation systems, diffusion models excel in generative capabilities, effectively capturing underlying data distributions and learning high-quality representations to accomplish diverse generative tasks. Their flexible architectures also allow seamless integration with other models. These advantages enable diffusion models to serve three primary roles in recommendation systems: (1) Data Augmentation \cite{wu2024diffusion, you2024context} and Representation Augmentation \cite{wu2019neural, wu2020diffnet++, chen2024g}; (2) Direct User Preference Estimation \cite{walker2022recommendation}; (3) Personalized Content Generation \cite{xu2024diffusion}. For specific technical details, readers may refer to the seminal article \cite{lin2024survey}, which provides a comprehensive overview of diffusion-based approaches in recommendation systems.

Overall, diffusion models constitute a pivotal component of generative recommendation systems and are poised to advance the evolution of next-generation recommendation systems.

\subsection{Variational Autoencoders}
VAE is a generative model based on a probabilistic framework, consisting of an encoder and a decoder. The \emph{encoder} maps input data to a distribution over latent variables, while the \emph{decoder} reconstructs the input data from the latent variables to generate recommendation results. 

Similar to GANs, VAEs can capture the latent distribution of input data and encode it into a lower-dimensional latent space, generating data samples similar to the input. In particular, VAEs are commonly used in collaborative filtering techniques. Compared to GANs, VAEs offer higher training efficiency but produce lower-quality generated data. As a result, some researchers \cite{zhou2023vcgan} have combined VAEs with GANs to improve recommendation systems, leveraging the strengths of both models.

In recent years, generative models have gained increasing attention, and recommendation methods based on VAEs have become more prevalent. The paper \cite{liang2024survey} is the first to systematically investigate the development of VAEs in the recommendation domain. This paper summarizes four core characteristics of VAEs: (1) Encoding capability; %Extracts latent representations of users and items through the encoder, enabling multimodal data fusion. 
(2) Generative capability;
%Generates user preference distributions, mitigating the cold-start problem and even generating new content. 
(3) Bayesian properties;
%Handles uncertainty through variational inference, enhancing model robustness.
(4) Flexible internal structure.
%Supports customizable encoders, decoders, prior distributions, and objective functions.
The paper also summarizes the applications of VAEs in recommendation systems, which can be categorized into \emph{Direct recommendation generation}, %Learns latent representations from user-item interaction data to predict user preferences; or integrates sequential data (e.g., user behavior sequences) by leveraging recurrent neural networks (RNNs) to enhance temporal modeling.  
\emph{Auxiliary information fusion}
%: Extracts latent representations from multimodal data such as text and images, integrating them with collaborative filtering; or employs conditional VAEs (CVAEs), using user profiles and item tags as conditional inputs.  
and \emph{Optimization strategy expansion}.
%: Improves prior distributions (e.g., user-dependent priors, mixture priors) to enhance model expressiveness or integrates generative adversarial networks (GANs) to increase generative diversity.

% The survey provides a detailed investigation and analysis of VAE-based recommendation algorithms, concluding that these algorithms can handle sparse data, facilitate multimodal data fusion, and possess Bayesian inference and generative capabilities. 

% This implies that VAEs are likewise an indispensable component in the landscape of generative recommendation systems.
These studies suggest that VAEs are also a vital component within the landscape of generative recommendation systems. % However, they encounter several challenges, including posterior collapse, limited diversity in generated outputs, high computational costs, and sensitivity to hyperparameter settings. \textcolor{blue}{Why discusses challenges here? I see other three kinds of models only discuss advantages.}

%However, they also face challenges such as posterior collapse, where latent variables fail to effectively capture data information, limited diversity in generated results, as well as high computational complexity and sensitivity to hyperparameters.

\subsection{Large Language Models}
%\cite{liu2023pre}
LLMs are transformer-based models with a vast number of parameters, trained on massive datasets using self-supervised or semi-supervised learning techniques. %The core strength of LLMs lies in their ability to extract high-quality textual feature representations and encode extensive external knowledge, exhibiting strong natural language comprehension and text generation capabilities. As a result, 

Incorporating LLMs into recommendation systems is expected to enhance their ability to capture contextual information and better understand textual data, such as user queries, item descriptions, user comments and feedback, to extract high-quality feature representations. As a general-purpose model, LLMs also have other notable advantages. (1) It can be applied to various recommendation tasks simply by using different prompt instructions, showing strong generalization ability. (2) By leveraging prompt strategies such as chain-of-thought prompting, it can perform multi-step reasoning, exhibiting powerful reasoning capability. (3) Through zero-shot or few-shot recommendation, it can easily address the sparsity problem of historical interaction data. Due to these advantages of LLMs, recommendation systems based on LLMs play a crucial role in the field of generative recommendation. 

Numerous studies have already emerged exploring the application of LLMs in recommendation systems \cite{wu2024survey, li2023large, zhao2023recommender, lin2025can, harte2023leveraging, friedman2023leveraging, chu2023leveraging}. 
We can note that current methodologies for integrating LLMs into recommendation systems commonly include pre-training, fine-tuning, and prompt engineering \cite{liu2023pre}. \emph{Pre-training} constitutes a critical phase in developing LLMs, involving training on massive datasets and corpora to equip them with broad world knowledge and reasoning and generative capabilities \cite{wu2020ptum, cui2022m6}. While pre-training enables LLMs to acquire extensive knowledge and capabilities, the associated computational costs are substantial. Therefore, many studies focus on fine-tuning pre-trained LLMs to enhance their performance for specific applications. \emph{Fine-tuning} refers to the essential process of adapting pre-trained LLMs to downstream tasks, such as recommendation systems. Through fine-tuning, LLMs gain domain-specific knowledge ($e.g.$, user preferences, item attributes) and demonstrate task-specific expertise ($e.g.$, personalized ranking, diversity-aware recommendation). This approach balances computational efficiency with performance optimization, making it a cornerstone for deploying LLMs in practical recommendation scenarios. Relevant studies on improving the capability of LLMs for recommendation systems through fine-tuning include \cite{zheng2023generative, friedman2023leveraging, kim2021intent, mao2023unitrec, bao2023tallrec}. \emph{Prompt engineering} stands out as the most expedient and resource-efficient approach. It involves adapting LLMs to downstream tasks through task-specific prompts, eliminating the need for extensive parameter updates. Widely adopted prompt-based techniques include In-context learning (ICL), Chain-of-thought (CoT) and Prompt tuning. ICL leverages examples within the input prompt to guide LLMs in generating context-aware recommendations. CoT prompting explicitly structures prompts to simulate reasoning steps. Prompt tuning optimizes lightweight, learnable prompt embeddings while keeping the LLM parameters frozen. Recent studies include \cite{mann2020language, wei2022chain, liu2023chatgpt, gao2020making}.

By aligning recommendation systems with LLMs, these three techniques enable LLM-based recommendation systems to achieve enhanced performance, positioning them as a pivotal research focus in the field.

\section{Generative Multi-objective recommendation systems}

\begin{figure*}[h]
    \centering
    \includegraphics[width=0.6\textwidth]{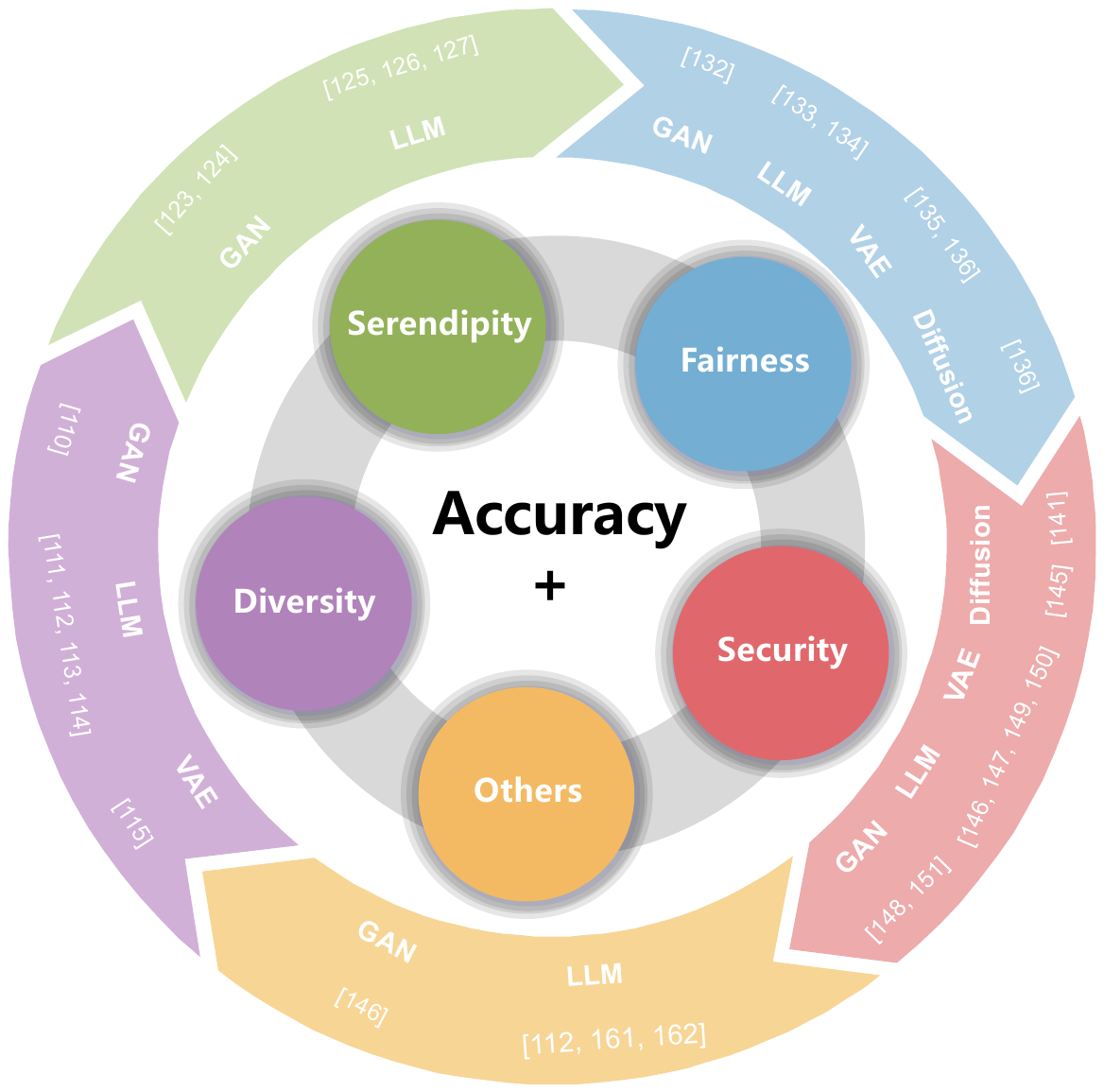}
    \caption{An overview of existing studies in multi-objective generative recommendation systems}
\label{overview}
\end{figure*}

With the widespread applications of recommendation systems in various fields, single-objective strategies are increasingly insufficient to address the complex and often conflicting needs of users and system stakeholders. The rise of generative technology provides new possibilities for multi-objective optimization. In-depth research on the optimization methods of generative MORS is of great significance for improving the overall performance of the system and its practical applications. In this section, we will discuss in detail the application of generative technology in MORS, along with the evaluation methods used for various objectives. Fig \ref{overview} depicts an overview of existing studies.

w
\subsection{Diversity}

Among the critical goals of recommendation systems, diversity plays a pivotal role in improving user experiences. 
% The integration of diversity within recommendations offers several benefits. It increases relevance by matching suggestions to specific tastes, while simultaneously broadening the user's horizons through varied content. This dual benefit contributes to greater user satisfaction and engagement \cite{sah2024unveiling}.
Pure accuracy-targeted methods may lead to the echo chamber or filter bubble effects \cite{ge2020understanding}, trapping users in a small subset of familiar items without exploring the vast majority of others. To break the filter bubble, diversification in recommendation systems has received increasing attention. Through an online A/B test, research \cite{huang2021sliding} shows that the number of user engagements and the average time spent greatly benefit from diversifying the recommendation systems. Diversified recommendation targets increase the dissimilarity among recommended items to capture users’ varied interests. Nevertheless, optimizing diversity alone often leads to decreases in accuracy. Accuracy and diversity dilemma \cite{ziegler2005improving} reflects such a trade-off. Therefore, diversified recommendation systems aim to increase diversity with minimal costs on accuracy \cite{ashkan2015optimal, chen2018fast, zheng2021dgcn}.

% An essential aspect of achieving diversity in recommendation systems lies in accurately understanding user preferences and personality traits, which requires a reliable assessment of user characteristics. Several approaches have been developed to infer user traits, each with its strengths and limitations. For example, self-report surveys and standardized personality tests rely on direct user input, but may suffer from biases or logistical challenges. Observational data and social media data offer alternative insights through indirect means, yet they may lack comprehensive or representative coverage of a user's personality. 
Recently, LLMs have emerged as a novel approach, leveraging users' text-based interactions to infer traits. These methods form the foundation for generating personalized recommendations, but their limitations also affect the diversity of suggestions provided to users. For instance, biases in user data or inference models may skew recommendations towards specific preferences, potentially reducing exposure to diverse content. Addressing these challenges requires integrating robust data sources and refining inference techniques to ensure that both personalization and diversity are achieved harmoniously.

\subsubsection{Definition}
In most existing studies, diversity is defined based on the redundancy or similarity among the recommended items, where the detailed difference might come from the categories \cite{vargas2014coverage} or the distance in item embedding space \cite{vargas2011intent, shi2012adaptive}. Items are naturally different from each other, and the level of diversity is different across items. For example, item diversity might come from categories at a high level or from item features at a detailed level. Item features can be explicit ($i.e.$, intrinsic item features) or implicit ($i.e.$, from learned embeddings considering both item features and other users’ history interactions). Therefore, there are different metrics that focus on varying granularities and aspects of item diversity.

\subsubsection{Models}
In this subsection, we review existing models for improving diversity in the field of generative recommendations. 

In the paper \cite{wu2019pd}, a novel recommendation model using the GAN framework is built, called Personalized Diversity-promoting GAN (PD-GAN), which consists of a generative network (generator) and a discriminative network (discriminator) contesting with each other. To generate recommendations that are as diverse and relevant as the ground-truth, this work considers to diversify recommendation results for users with broad interests, while avoiding the blind pursuit of diversity for users with focused preferences.

In particular, Xu $et$ $al.$ \cite{10823618} show that LLM has significant advantages in the field of personalized recommendation, which can improve user experience and promote platform sales growth, and provides strong theoretical and practical support for personalized recommendation technology in e-commerce. It adopts a hybrid model structure to calculate the final recommendation score, combining LLM (Pretrained BERT) with existing collaborative filtering and content-based recommendation (CBF) algorithms. The model they proposed achieved substantial increases in both accuracy and diversity.

DLCRec \cite{chen2024dlcrec} poses a novel framework designed to enable fine-grained control over diversity in LLM-based recommendations. Unlike traditional methods, DLCRec adopts a fine-grained task decomposition strategy, breaking down the recommendation process into three sequential sub-tasks: genre prediction, genre filling, and item prediction. All three sub-tasks were completed using large models. These sub-tasks are trained independently and inferred sequentially according to user-defined control numbers, ensuring more precise control over diversity. %They use two control metrics: Cov@K and MAE\_Cov@K, to evaluate the model’s ability to meet the diversity requirements. Cov@K measures the coverage of genres in the recommended list, while MAE\_Cov@K calculates the mean average error between the actual and desired genre coverage.%

While D3Rec \cite{han2024controlling} shares DLCRec's fundamental objective of controlling accuracy and diversity in recommendations, it differentiates itself through implementation of a diffusion-based paradigm. In the forward process, D3Rec eliminates category preferences lurking within user interactions by adding noises. Then, in the reverse process, D3Rec generates recommendations through denoising steps while reflecting the targeted category preference. The system can enhance diversity by adjusting the targeted category preference smoothly, making it applicable to various real-world scenarios involving different desired category distributions.

In \cite{carraro2024enhancing}, LLMs are prompted to generate a diverse ranking from a candidate ranking using various prompt templates with different re-ranking instructions in a zero-shot fashion. They conduct comprehensive experiments testing state-of-the-art LLMs from the GPT and Llama families. They compare their re-ranking capabilities with random re-ranking and various traditional re-ranking methods from the literature. Their findings suggest that the trade-offs (in terms of performance and costs, among others) of LLM-based re-rankers are superior to those of random re-rankers but, as yet, inferior to those of traditional re-rankers. %To measure recommendation diversity, they employ $\alpha$-NDCG, Intra-List Diversity (ILD), Expected Intra-List Diversity (EILD), Subtopic-Recall (SRecall) and relevancy-aware SRecall (rSRecall).%

Proposed by \cite{izawa2025diversifying}, a tailored model for furniture recommendation builds upon the Recommendation Variational Autoencoder (RecVAE), known for its effectiveness and ability to overcome overfitting by linking user feedback with user representation. RecVAE’s implicit feedback reliance induces popularity bias, risking filter bubble effects. Contrasting prior user profiles from personal information and item text, it proposes image-derived profiles. They posit that accurate and diversified furniture recommendation requires integrating both textual reviews and visual data. To optimize preference modeling, they employ a Conditional VAE (CVAE) architecture, where the encoder and decoder are conditioned on style-aware user profiles trained to capture predefined aesthetic preferences.

However, achieving diversity poses significant challenges, including mitigating biases in data and algorithms, and addressing ethical concerns such as user autonomy and potential manipulation.

\subsubsection{Evaluation metrics}
In this subsection, we collect common metrics used in diversity measurement. Some papers propose unique diversity metrics, which are beyond the scope of our discussion.

\begin{itemize}
    \item $\alpha$-\textit{NDCG}. By plugging redundancy-aware gain in NDCG, we get the metric known as $\alpha$-NDCG. Introduced by \cite{clarke2008novelty}, it can quantify the trade-off between diversity and relevance with a single metric. It is calculated as:
    \begin{align}
    & r_{i, k-1} = \sum_{j=0}^{k-1} J(d_j, n_i) \nonumber, \\
    & G(k) = \sum_{i=1}^m J(d_j, n_i) \, (1 - \alpha)^{r_{i, k-1}} \nonumber, \\
    & \alpha\text{-}\text{NDCG} = \frac{1}{\alpha\text{-}\text{IDCG}} \sum_{k=1}^n \frac{G(k)}{\log_2 (k+2)},
    \end{align}
    where $n$ is the number of documents in the ranked list, $J(d_j, n_i)$ is the binary relevance judgment of document $d_j$ with respect to subtopic $n_i$, $m$ is the total number of subtopic and $\alpha$-IDCG is the normalization factor set as the maximum possible value of $\alpha$-DCG for an ideal ranking.

    \item \textit{Intra-List Diversity (ILD).} ILD measures the average pairwise dissimilarity between items in a recommendation list \cite{zhang2008avoiding}. It is calculated as:
    \begin{align}
    \text{ILD} = \frac{2}{|L|(|L|-1)} \sum_{i \in L} \sum_{j \in L, j \neq i} (1 - {d(i, j)}),
    \end{align}
    where $|L|$ is the number of items in the list, $d(i, j)$ is the distance or dissimilarity between items $i$ and $j$. The variants differs in the way of obtaining item embeddings and the specific distance function.

    \item \textit{Entropy.} Entropy quantifies the diversity of the recommendation distribution \cite{bellogin2010study}. Higher entropy indicates a more uniform distribution of recommended items.
    \begin{align}
    \text{Entropy} = - \sum_{i \in I} p(i) \log p(i),
    \end{align}
    where $p(i)$ is the probability of item $i$ being recommended, typically estimated as the frequency of item $i$ in all recommendation lists.

    \item \textit{Coverage.} In most cases, it measures the proportion of unique items recommended across all users. It is also called aggregate diversity \cite{Vargas2015}. It is calculated as:
    \begin{align}
    \text{Coverage} = {|\bigcup_{u\in U} L_u|},
    \end{align}
    where ${\bigcup_{u\in U} L_u}$ represents the union of all items recommended to users in the set $U$ and $|I|$ is the total number of items in the catalog. However, in some cases, it measures the coverage of genres in the recommended list, which is also denoted as Category Coverage (CC) \cite{puthiya2016coverage}.
\end{itemize}

\subsection{Serendipity}

Research on serendipity-oriented recommendation systems helps address issues in traditional recommendation systems, such as the cold start problem and filter bubbles. In model construction, commonly used serendipity-oriented recommendation algorithms can be categorized into three types. The first type consists of re-ranking algorithms. For example, the paper \cite{tokutake2023serendipity} enhances an existing accuracy-focused recommendation system by employing a re-ranking algorithm and a time series prediction model to dynamically optimize unexpectedness in recommendations. Similarly, the paper \cite{park2022psr} re-ranks recommendations by combining user-provided unrelated tags, generating unexpected creative item combinations. The second type of method modifies accuracy-focused recommendation models to incorporate serendipity, though such modifications are only applicable to certain algorithms. In addition, the third type involves proposing new techniques that do not rely on existing accuracy-oriented recommendation models.

The advancement of generative techniques in recommendation systems introduces new possibilities, potentially achieving a better balance between accuracy and serendipity. By moving beyond accuracy-centric paradigms, such approaches promise to foster more diverse, engaging, and user-centric recommendation experiences, while mitigating risks such as content homogeneity and filter bubbles.%Hence, in this section, we will discuss current research on serendipity in the field of generative recommendation, including definition, model architectures, and relevant evaluation metrics.

\subsubsection{Definition}

The concept of serendipity is highly subjective, leading to variations in its definition across individuals. In the field of generative recommendation, research on serendipity remains limited. Existing studies generally define serendipity in a manner similar to its definition in traditional recommendation systems, which is considering serendipity as a composition of several sub-objectives. For example, in paper \cite{xu2023gs}, serendipity is defined as low initial interest but high user satisfaction. And in \cite{hassan2024gczrec}, serendipity is defined as "unexpected and diverse", while in \cite{fu2024art}, the authors define serendipity as comprising unexpectedness and relevance, excluding diversity and novelty from its formulation. %Then in \cite{xi2025bursting}, serendipity is similarly defined as a combination of unexpectedness and relevance and an industrial definition for online shopping platforms is also proposed. 
Similarly, \cite{xi2025bursting} defines serendipity as a combination of unexpectedness and relevance, and also proposes an industrial interpretation for online shopping platforms.
Additionally, we also note that some studies do not decompose serendipity into multiple sub-objectives, such as in \cite{tokutake2024can}, serendipity is treated as a holistic measure for evaluation.

\subsubsection{Models}
Here, we discuss the existing studies on serendipity in the field of generative recommendation.

In the paper \cite{hassan2024gczrec}, the GCZRec framework is proposed, which addresses the utilization of collaborative signals and diversity challenges in cold-start recommendation systems combining GANs and zero-shot learning. Specifically, the framework employs a zero-shot classifier to predict labels for cold-start news items or users, which are then fed into two generators (genN and genU) to produce interest score vectors. These vectors are subsequently synthesized to generate final recommendations. In this study, serendipity is defined as "unexpected and diverse". Experimental results demonstrate that the framework outperforms baseline models on accuracy metrics such as NDCG. Additionally, multi-generational recommendation experiments reveal a progressive increase in the proportion of novel items in the recommendations, indicating that the framework achieves both enhanced ranking quality and improved serendipity. The key to optimizing these dual objectives lies in the implicit exploitation of collaborative signals and multi-label predictions for cold-start users or items.

Another paper \cite{xu2023gs} proposes a new serendipity-oriented recommendation system framework called GS2-RS, using GANs to address the cold start and filter bubble problems. In this context, the filter bubble problem is transformed into the challenge of enhancing the diversity of recommendation results. The paper defines serendipity as low initial interest but high user satisfaction. It utilizes twin conditional GANs to extract fine-grained preferences from the input user-item interaction matrix, generating reliable virtual neighbor preferences. A gating mechanism is used to generate each user's self-serendipity preferences. Items that are unlikely to be of interest are marked as 0, thereby producing an enhanced matrix that is then fed into an existing recommendation system.

In \cite{fu2024art}, the study explores the use of LLMs for serendipity-based recommendations by designing different prompts to guide the model in generating unexpected recommendations and determining whether a given item is surprising for a specific user. Three types of prompts are proposed: \emph{discrete prompts} based on natural language design, \emph{continuous prompts} incorporating learnable tokens, and \emph{hybrid prompts} that combine natural language with virtual tokens. All three types of prompts include two paradigms: direct prompting and indirect prompting. Direct prompting treats serendipity as a whole and directly predicts it, while indirect prompting decomposes serendipity into two objectives, unexpectedness, and relevance, for separate predictions. The experimental results ultimately demonstrate the effectiveness of LLMs in serendipity recommendation and show that decomposing serendipity into unexpectedness and relevance improves the recommendation performance of LLMs. 

In \cite{tokutake2024can}, the authors investigate whether LLMs can evaluate users' serendipitous reactions. The task is framed as a binary classification problem: given a user’s rating history and recommendations generated by existing systems, the LLM predicts either “yes” or “no” to indicate whether the recommendation is serendipitous. By comparing the model predictions with the human assessment, the study finds that while the alignment between the serendipitous evaluations made by LLMs and those made by humans is not particularly high, it still outperforms the baseline models. Furthermore, this study reveals that both insufficient and excessive user rating histories negatively impact model performance, and the interpretability of high-quality output remains a challenge.

Building on the previous two studies \cite{fu2024art, tokutake2024can}, in \cite{xi2025bursting}, the SERAL model is introduced to address the following challenges: the inconsistency between LLM-based serendipity judgments and human evaluations, the limited user history length that LLMs can process, and the high latency of online inference. The model consists of three main components. \emph{Cognition Profile Generation} mimics the human cognitive formation process by compressing user information and long behavioral sequences into a concise, multi-level cognitive profile, alleviating the difficulty of LLMs in handling long sequences. \emph{SerenGPT Alignment} employs a preference alignment algorithm to align serendipity judgments with human preferences using enriched training data. \emph{Nearline Adaptation} caches high-quality candidate recommendations generated by SerenGPT and integrates them into traditional multi-stage recommendation pipelines, reducing inference latency. Through deployment on Taobao, the study demonstrates that LLMs can help break recommendation filter bubbles and enhance user satisfaction.

It is also worth noting that in the studies above, serendipity itself encompasses multiple goals. Therefore, developing serendipity-driven recommendation systems can be seen as achieving multi-objective optimization. Compared to accuracy-focused recommendation systems, this approach provides users with a more comprehensive and enriching experience. Overall, these studies effectively validate the potential of generative AI in serendipity recommendation. 

However, current research remains in its early stages. Further research is encouraged to advance this field, particularly through cross-disciplinary collaboration and innovative methodologies that deepen the theoretical foundations and practical applications of serendipity-aware recommendation systems.

\subsubsection{Evaluation metrics}
The difficulty in defining serendipity also makes its evaluation a challenging task in research. In this subsection, we collect common metrics used in serendipity measurement for generative recommendation. 

\begin{itemize}
    \item \textit{$\text{NDCG}_{seren}@k$}. This metric is derived from the metric NDCG \cite{fu2024art,xi2025bursting}. The formula is defined as follows:
    \begin{align}
     \text{NDCG}_{seren}@k = \sum^k_{i=1}\frac{serendipity\ score\ (1\ or\ 0)}{log_2(i+1)}.
    \end{align}

    \item \textit{$\text{HR}_{seren}@k$}. It is a metric to measure the proportion of times the serendipity item is retrieved in the top-$k$ position \cite{fu2024art,xi2025bursting}. It is calculated as:
    \begin{align}
    \text{HR}@k = \sum^k_{i=1}serendipity\ score\ (1\ or\ 0).
    \end{align}
    
    \item \textit{$\text{MAP}_{seren}@k$}. This is a serendipity-version of MAP (Mean Average Precision) \cite{fu2024art, xi2025bursting}. It is necessary to first define $\text{AP}_{seren}@k$, which is a serendipity version of Average Precision. The formulas are as follows:
    \begin{align}
    \begin{split}
         \text{AP}_{seren}@k =&\frac{1}{N_{seren}}\sum^k_{i=1}\text{HR}@k 
     \\ &\times serendipity\ score\ (0\ or\ 1),
    \end{split}    
    \end{align} 
    \begin{align}
     \text{MAP}_{seren}= \frac{1}{Q}\sum^Q_{i=1} \text{AP}_{seren}(q),
     \end{align} 
     where $Q$ denotes the total number of users.

    \item \textit{Diversity (DI)}. This is a direct metric to evaluate the information entropy gain of a recommendation \cite{xu2023gs}. The formula is as follows:
    \begin{align}
        \text{DI}@k = \frac{category(I^{rec})}{num(I^{rec})},
    \end{align}
    where, $I^{rec}$ means the recommendation item list, including k items.
    
    \item \textit{Serendipity (SE)}. This is a metric to represent how many high-satisfaction but low-interest items are recommended \cite{xu2023gs}. The formula is as follows:
    \begin{align}
        \text{SE}@k = \frac{num(I_{rec}\cap I_{real}\cap (I_{sa} -I_{in}))}{num(I_{real})},
    \end{align}
    where, $I_{rec}$ means the same above, and $I_{real}$ means the ground-truth of the test set. $I_{sa}$ and $I_{in}$ mean the high-satisfaction and the low-interest item list, respectively.

\end{itemize}

Notably, beyond quantitative metrics, serendipity has also been measured in research through survey-based methods, where user feedback is systematically collected to evaluate the perceived novelty and unexpected relevance of recommendations, as shown in \cite{guan2024review}.

\subsection{Fairness}

The rise of LLMs has significantly enhanced the ability of recommendation systems to provide diversified suggestions. %However, alongside the advances in diversity, fairness has emerged as another critical concern. LLMs may inherit and amplify biases present in the training data, leading to unfair or discriminatory recommendations.
However, this progress has also intensified concerns about fairness. LLMs can inherit and even amplify biases embedded in their training data, potentially leading to unfair or discriminatory recommendations. The goal of fairness research is to mitigate such biases introduced during training. For example, \cite{sah2024unveiling} explores the combination of personal profiles of users with LLMs to create customized recommendations, thus reducing bias, better aligning with user needs and enhancing the overall user experience. In this section, we systematically analyze the fairness issues in recommendation systems based on the main perspectives proposed in \cite{hua2023tutorial}, including the definition, evaluation, and models.

\subsubsection{Definition}

Generative AI, particularly LLMs, has demonstrated significant potential in information management systems, capable of generating various forms of content such as text, images, and audio. However, as these technologies become more widely adopted, the issue of fairness has emerged as a critical concern. Fairness in Generative AI can be defined as the model's ability to avoid systematic errors or biases when processing information, ensuring that its outputs do not exhibit unjust preferences or discrimination against specific groups. Specifically, Generative AI models learn patterns from large-scale datasets during training, which often reflect societal biases and inequalities present in the real world. If these biases are inherited without correction, the generated content may reinforce stereotypes related to gender, race, culture, and other dimensions, leading to unfair decisions and outcomes. Therefore, the core objective of fairness in Generative AI is to ensure that the model's outputs are equitable for all user groups, avoiding adverse impacts on certain demographics due to biases in data or algorithms \cite{wei2025addressing}.

\subsubsection{Models}

Here, we discuss existing research advances related to fairness in the field of generative recommendations.

Zhang \emph{et al.} \cite{zhang2024gan} propose a fairness-aware recommendation algorithm based on GAN, called Prior Fair GAN (PFGAN). Its core idea is to improve the fairness of the recommendation system while maintaining high recommendation accuracy by balancing the capabilities of the generator and the discriminator. Specifically, PFGAN introduces Accumulated Individual Exposure Disparity (AIED) as a regularization term in the objective function of the generator to ensure that the recommendation results not only have high user utility, but also reduce the exposure difference of the overall data. Therefore, PFGAN can dynamically adjust the capabilities of the generator and the discriminator during the training process to avoid the training crash problem caused by the imbalance of the capabilities of the two.

Sakib \emph{et al.} \cite{sakib2024challenging} classify the types of recommended content, used the GPT-3.5 model to classify the recommended content, and calculated the distribution of different types in different groups. Compared with traditional generative techniques such as GAN, the application of GPT-3.5 in type classification tasks focuses more on semantic understanding and label generation rather than generating new data samples. However, this application still reflects the core characteristics of generative technology, which is to generate meaningful results through the model.

Unlike using LLM to complete content classification, the generative recommendation system uses the semantic understanding and generation capabilities of the generative model to handle complex user preferences and contextual information. The core technology of work~\cite{deldjoo2024understanding} is the generative language model based on ChatGPT for generating recommended content. Through natural language processing capabilities, it can understand the user's historical interaction data, such as movie ratings and music playback records, and generate a recommendation list. Through prompt design and system role embedding, generative technology can flexibly adjust the behavior of the recommendation system and balance multiple goals such as accuracy and fairness. For example, accuracy-oriented prompts can improve fairness to a certain extent while maintaining high accuracy. In addition, the generative model can quickly adapt to new tasks and data through in-context learning without additional training. %Zero-shot learning performs well in accuracy and coverage, while few-shot learning can improve accuracy in some cases, such as when age group information is included.

VAEs have also shown promise for fairness optimization due to their probabilistic latent space. Borges \emph{et al.} \cite{borges2019enhancing} innovatively introduce noise distributions, such as uniform noise and Gaussian noise, during the test phase. By disturbing the sampling process of latent variables, the recommendation results of the same input produce reasonable variations, and long-term fairness is achieved directly through randomness within the model. %Experiments show that the introduction of noise can significantly reduce unfairness, but the accuracy will be slightly reduced.

%Paper \cite{lilienthal2024synthetic} mentions the Fair Diffusion recommendation Module (FairDiff), which uses conditional variational autoencoder (CVAE) and diffusion model to generate group-aware negative samples. 
Lilienthal \emph{et al.} \cite{lilienthal2024synthetic} introduce FairDiff, a fairness-aware module combining Conditional VAE and diffusion models to generate group-aware negative samples. Its core idea is to embed user attributes into the latent space, and then generate negative preferences related to group labels through the diffusion process to improve the fairness of the recommendation system. 

However, there are limitations in current research, such as the difficulty in quantifying fairness indicators and the trade-off dilemma in balancing fairness and recommendation accuracy. In addition, generative models lead to a lack of transparency and explainability in their decision-making process, further increasing the difficulty of fairness evaluation and optimization.

\subsubsection{Evaluation metrics}

The following metrics can help us evaluate the fairness of the recommendation system and ensure that the algorithm does not produce bias in its performance between different groups or individuals.

In the context of LLM-based recommendation systems, fairness can be evaluated through three key dimensions: individual fairness, group fairness, and counterfactual fairness \cite{sah2024unveiling}.

\textbf{Individual fairness} means that users with similar characteristics and preferences should receive similar recommendations.

\begin{itemize}
    \item \emph{Sensitive-to-Neutral-Similarity-Range (SNSR) and Sensitive-to-Neutral-Similarity-Variance (SNSV).} In \cite{zhang2023chatgpt} and \cite{deldjoo2024cfairllm}, SNSR and SNSV are proposed to quantify the degree of unfairness. The higher the value, the higher the degree of unfairness. For top-$k$ recommendation, it is calculated as:

    \begin{equation}
    \small
    \text{SNSR}@k = \underset{a \in \mathcal{A}}{\text{max}} \overline{Sim} (a) -  \underset{a \in \mathcal{A}}{\text{min}} \overline{Sim} (a);
    \end{equation}

    \begin{equation}
    \small
    \text{SNSV}@k = \sqrt{ \frac{1}{|\mathcal{A}|} \underset{a \in \mathcal{A}}{\sum} (\overline{Sim} (a)- \frac{1}{|\mathcal{A}|} \underset{a' \in \mathcal{A}}{\sum} \overline{Sim} (a'))^{2}}.
    \end{equation}

    Here, $|\mathcal{A}|$ represents the number of all possible values in the sensitive attribute under study. SNSR measures the difference between the similarity of the most advantaged group and the most disadvantaged group, while SNSV calculates all possible variances of the sensitive attribute $\mathcal{A}$.
\end{itemize}

\textbf{Group fairness} focuses on ensuring that different demographic groups, such as gender, age, and race, are not subjected to systematic biases in the recommendations provided by the system, which seeks to mitigate the risk of discriminatory outcomes that could arise from underlying data or model biases.

\begin{itemize}
    \item \emph{Ranking-based Statistical Parity (RSP) and Ranking-based Equal Opportunity (REO).} Zhu $et$ $al.$ \cite{zhu2021fairness} propose RSP, which is calculated as:

    \begin{align}
    &P_i = P(\text{R}@k \mid g = g_i) \nonumber \\
    &\text{RSP}@k = \frac{{std}(P_1, P_2, \dots, P_A)}{{mean}(P_1, P_2, \dots, P_A)}.
    \end{align}

    And REO is calculated as:

    \begin{align}
    &P_i' = P(\text{R}@k \mid g = g_i, y = 1) \nonumber \\
    &\text{REO}@k = \frac{{std}(P_1', P_2', \dots, P_A')}{{mean}(P_1', P_2', \dots, P_A')}.
    \end{align}

    Here, $P_i$ represents the probability of items in group $g_a$ being ranked in top-$k$; $P_i'$ represents the probability he probability of being ranked in top-$k$ when the user likes the item, where $y=1$ represents items are liked by users; ${std}$ denotes the standard deviation, and \({mean}\) denotes the average. For both, lower values indicate fairer recommendations.
    
    % \item \emph{Equality of Representation.} In addition to these two more popular group fairness metrics, Fairwalk proposes a notion called Equality of Representation \cite{vcutura2024fairness}, which defines bias through two variants: network level and user level \cite{rahman2019fairwalk}, to promote equal recommendations for all groups in OSN graph embeddings.
    \item \emph{Equality of Representation.} In addition to these two more popular group fairness metrics, Equality of Representation~\cite{vcutura2024fairness} is proposed, which defines bias through two variants: network level and user level \cite{rahman2019fairwalk}, to promote equal recommendations for all groups in OSN graph embeddings.

    \item \emph{Fairness Score.} Rao $et$ $al.$ \cite{rao2023can} propose the “Fairness Score” indicator. When only analyzing the fairness of LLM evaluation for different genders, it is calculated as:
    \begin{equation}
    s_r = \frac{\alpha}{\alpha + D_E(\overline{X'}, \overline{X})},
    \end{equation}
    where $X’$ and $X$ represent the average test results for males and females.
\end{itemize}

\textbf{Counterfactual fairness} means that recommendations remain fair even if user characteristics change. This dimension explores the robustness of fairness under dynamic conditions, ensuring that recommendations remain equitable regardless of user attribute variations.

\begin{itemize}
    \item \emph{Gini Index.} The study of \cite{deldjoo2024understanding} introduces the Gini Index as a metric to quantify the inequality in the distribution of recommended items. A lower Gini Index indicates a more equitable distribution among items. It is calculated as:
    \begin{equation}
    \text{Gini} = \frac{\sum_{i=1}^{n} (2i-n-1) x_i}{n \sum_{i=1}^{n} x_i},
    \end{equation}
    where $x_i$ represents the number of times item $i$ was recommended and $n$ represents the total number of items.

    \item \emph{Individual Exposure Disparity (IED).} Based on the Gini Index, IED is proposed~\cite{zhang2024gan} to measure the exposure fairness of a single recommendation result, reflecting the difference in exposure of each item in the recommendation list. It is calculated as:

    \begin{equation}
    \text{IED} (\mathcal{L}) = \frac{\sum_{v, v' \in \mathcal{I}} | Exp (v|\mathcal{L}) - Exp (v|\mathcal{L})|}{2n \sum_{v'' \in \mathcal{I}} Exp (v''|\mathcal{L})},
    \end{equation}

    where $Exp (v|\mathcal{L})$ represents the exposure of item $v$ in the recommendation list $\mathcal{L}$. The lower the IED value, the fairer the exposure distribution of the recommendation result.

    \item \emph{AIED.} IED only focuses on the fairness of a single recommendation result, ignoring the impact of the recommendation system on the overall data distribution in the long-term operation. For this reason,  AIED is further proposed \cite{zhang2024gan}, which combines the prior fairness information in the dataset to measure the impact of the recommendation result on the overall data fairness. It is calculated as:
    
    \begin{equation}
    \text{AIED} (k) = \text{IED} (Exp(\mathcal{L}^k) + Exp(M)),
    \end{equation}
    
    where $M$ represents the user's historical interaction data. By introducing AIED, we can not only focus on the fairness of a single recommendation result, but also consider the impact of the recommendation system on data distribution in the long run, so as to more comprehensively evaluate the fairness of the recommendation system.
\end{itemize}

Furthermore, fairness metrics for rankings aim at ensuring that different groups are treated fairly in the ranking when generating ranking results. Kliachkin \emph{et al.} \cite{kliachkin2024fairness1} give a variety of indicators for evaluating \textbf{ranking fairness}, including proportional fairness (P-Fairness) and infeasible index.

\begin{itemize}
    \item \emph{P-Fairness.} It requires that the number of individuals in each protected group in the prefix of each ranking meets a given ratio. In $(\vec{\alpha}, \vec{\beta})-k$ $fair$ $ranking$, it is calculated as:
    \begin{multline}
        \forall_{prefix} P: |P| \geq k, \forall i \in [g], \\
        \lfloor \alpha_i \cdot |P| \rfloor \geq |P \cap G_i| \geq \lceil \beta_i \cdot |P| \rceil,
    \end{multline}
    where $\alpha_i$ and $\beta_i$ are the upper and lower limit ratios of each protected group, and $G_i$ is the $i$-th protected group. In addition, in $(\vec{\alpha}, \vec{\beta})-weak$ $k-fair$ $ranking$, it is calculated as:
    \begin{equation}
        \forall i \in [g], \lfloor \alpha_i \cdot k \rfloor \geq |P \cap G_i| \geq \lceil \beta_i \cdot k \rceil
    \end{equation}
    This variant only requires proportional fairness in the first $k$ positions of the ranking.

    \item \emph{Infeasible Index.} It measures the percentage of positions in the ranking that do not meet P-Fairness. Based on $\text{LowerViol} (\pi)$, the number of positions in the ranking below the lower limit bound and $\text{UpperViol} (\pi)$, the number of positions in the ranking above the upper limit bound, infeasible index is called $\text{TwoSidedInfInd} (\pi)$ and it is calculated as:
    \begin{equation}
        \text{TwoSidedInfInd} (\pi) = \text{LowerViol} (\pi) + \text{UpperViol} (\pi)
    \end{equation}
    
\end{itemize}

\textbf{Statistical Parity Difference (SPD), Disparate Impact (DI), and Equal Opportunity Difference (EOD)} are common metrics used to quantify bias in LLM-based recommendations \cite{sakib2024challenging}. Consider a dataset $D = (X, Y, Z)$, where $X$ represents the training data, $Y$ represents the binary classification label, $Z$ is the sensitive attribute, and $\hat{Y}$ represents the predicted label, then these metrics can be defined as follows.

\begin{itemize}
    \item \emph{SPD.} It measures the difference in the probability of different groups getting favorable results. It is calculated as:
    
    \begin{equation}
      \text{SPD} = P(\hat{Y} = 1 | Z = Q) - P(\hat{Y} = 1 | Z = \overline{Q}).
    \end{equation}
    
    An SPD value of $0$ indicates perfect fairness, that is, the probability of different groups getting favorable results is equal.

    \item \emph{DI.} It measures the probability ratio of different groups to obtain favorable results. It is calculated as:
    
    \begin{equation}
      \text{DI} = \frac{(\hat{Y} = 1 | Z = Q)}{P(\hat{Y} = 1 | Z = \overline{Q})}.
    \end{equation}
    
    A DI value of $1$ indicates perfect fairness, meaning that the ratio of favorable results for both groups is equal.

    \item \emph{EOD.} It focuses on the difference in the probability of different groups obtaining favorable results when the true label is positive. It is calculated as:
    
    \begin{multline}
        \text{EOD}= P(\hat{Y} = 1 | Z = Q, Y = 1) \\
        - P(\hat{Y} = 1 | Z = \overline{Q}, Y = 1).
    \end{multline}
    
    An EOD value of $0$ indicates perfect fairness, that is, when the true label is positive, there is no difference in the probability of different groups obtaining favorable results.
\end{itemize}

\subsection{Security}

%recommendation systems rely on big data to shape user habits, but the sensitive nature of user data and its susceptibility to unauthorized access make ensuring its security a critical development direction. The article \cite{himeur2022latest} provides a thorough review of security and privacy issues in recommendation systems amidst the rapid advancements in AI and machine learning. It identifies key challenges related to trust, authentication, end-user privacy, malicious attacks, fairness, bias, filter bubbles, and ethical concern.With the rapid advancement of large language models (LLMs), there is an urgent need for new research on the security of LLM-based recommendation systems. Undoubtedly, incorporating LLMs into recommendation systems tightly links the security of such systems to that of LLMs themselves.
Security and privacy have become critical research objectives in the development of recommendation systems. With the growing integration of generative technologies into recommendation systems, leveraging these techniques to enhance system security has emerged as a prominent research trend. Notably, generative technologies can serve a dual purpose: they not only offer new opportunities to strengthen the security of recommendation systems but also present novel avenues for attacking them. In this section, we explore security challenges associated with generative recommendation, examining both the use of generative methods to improve system robustness and their potential to compromise system integrity. We argue that understanding how generative techniques can be exploited to attack recommendation systems is essential for building more secure and resilient recommendation platforms.

\subsubsection{Definition}
Recommendation systems heavily rely on user data, making them susceptible to a range of privacy and security issues, including data leakage, malicious attacks, trust and authentication vulnerabilities, fairness concerns, and ethical dilemmas \cite{himeur2022latest}. Data leakage can result in the misuse of personal information, while malicious attacks such as fake rating attacks, adversarial attacks, and data poisoning undermine the reliability of the system. Trust and authentication vulnerabilities may lead to identity spoofing or unauthorized access, thereby compromising the credibility of recommendations. Ethical concerns arise when recommendation outcomes are used to manipulate user behavior or propagate politically biased or harmful content \cite{gao2024brief}. Therefore, a secure recommendation system must possess the integrated capability to defend against malicious threats, protect user data privacy, maintain the trustworthiness of its recommendations, and uphold ethical standards through a combination of technologies, mechanisms, and strategies.

\subsubsection{Models}

Here, we review existing research related to security in the field of generative recommendations.

In general, the security of recommendation systems is studied from both attack and defense perspectives. Therefore, we maintain this methodological orientation in our study.

\textbf{Generative models for defense.} Paper \cite{lilienthal2024multi} proposes SDRM for generating synthetic user-item interaction data to supplement or replace original datasets. SDRM employs a VAE-based encoding mechanism to project user-item preferences into a latent Gaussian space. A diffusion model then operates on this latent representation, and the transformed data is subsequently reconstructed through the VAE decoder. Experimental results demonstrate that the model can enhance the accuracy of the recommendation system while preserving user privacy. In another paper \cite{zhang2024lorec}, a LoRec framework is proposed to mitigate adversarial impacts. On the one hand, LoRec dynamically adjusts the influence of training samples through a user weight calibration mechanism, reducing the risk of mistakenly penalizing legitimate users. On the other hand, by incorporating the open-world knowledge of LLMs, the model learns from a broad range of fraudulent patterns, avoiding over-reliance on specific attack assumptions. As a result, it enhances system security with minimal impact on recommendation accuracy. In addition, the paper \cite{saputra2024secure} proposes a framework that integrates LLMs, machine learning operations (MLOps), and security by design. It leverages the natural language understanding capabilities of LLMs to enhance recommendation accuracy, employs multi-stage security design to dynamically defend against security threats, and utilizes MLOps to enable real-time data processing and model updates, thereby improving the responsiveness and adaptability of the recommendation system. To address a significant challenge called prompt sensitivity in LLM-based recommendation systems, which refers to that it is highly susceptible to the influence of prompt words, \cite{li2024ganprompt} proposes GANPrompt. By integrating GANs generation techniques to produce diverse prompt, this paper enhances the model’s adaptability and stability to unseen prompts, improving the robustness of recommendation systems in complex and dynamic environments.

\textbf{Generative models for attack.} Considering that LLM-based recommendation systems rely heavily on textual content ($e.g.$, titles and descriptions), the paper \cite{zhang2024stealthy} is the first to propose a novel stealthy textual attack paradigm. In this paradigm, an attacker can significantly increase the exposure of target items by making subtle modifications to item text, such as keyword insertion or paraphrasing during the testing phase, without interfering with model training. Similarly, another paper \cite{ning2024cheatagent} introduces a novel attack framework, termed CheatAgent, which leverages the human-like capabilities of LLMs to develop LLM-based agents for attacking LLM-powered recommendation systems. CheatAgent first identifies the insertion position where minimal input modifications yield maximal impact. It then designs an LLM agent to generate adversarial perturbations and insert them at the target position. Finally, the agent employs prompt tuning techniques to iteratively refine the attack strategy based on feedback from the recommendation system. Experimental results demonstrate the effectiveness of this attack method. Paper \cite{liu2025novel} proposes the RWA-GAN model, which employs a GAN to generate realistic fake user profiles. The model supports multi-objective attacks, aiming to enhance attack efficiency, stealthiness and transferability, while also offering new insights for advancing defense research in recommendation systems. The model adopts a three-level generation architecture: the generator minimizes the KL divergence to make the distribution of false data close to the real data; the discriminator improves the discrimination ability through binary cross entropy loss; the attack trainer uses a substitute model to simulate the target system and optimizes the attack effect through a multi-target loss function. Experiments show that this method effectively balances the attack effect and concealment. 

It is also worth noting that the integration of LLMs into recommendation systems has introduced new security challenges. While numerous studies \cite{zhiheng2023safety, yao2024survey, chua2024ai} have called for deeper investigation in this area, concrete improvement strategies remain limited. Consequently, effectively leveraging generative technologies to enhance recommendation performance while ensuring system security remains a key direction for future research in the field.

\subsubsection{Evaluation metrics}
Attack and defense methods are typically evaluated by comparing the performance of the recommendation system before and after the intervention. A significant drop in performance following an attack indicates its effectiveness. When attacks involve injecting fake data, the higher the similarity between the fake and real data, the more stealthy and difficult to detect the attack becomes. While most studies follow this general evaluation framework to assess system security, the specific metrics used to measure recommendation performance vary. Commonly adopted metrics include Hit Rate (HR) and NDCG. In the following, we introduce several commonly used evaluation metrics for security.

\begin{itemize}
    \item \emph{Jaccard similarity.} Paper \cite{lilienthal2024multi} calculates the Jaccard similarity between synthetic data and original data, that is, quantifies the similarity by dividing the size of the intersection of the sets by the size of the union. It is calculated as:

    \begin{equation}
    J (x, \hat{x}) = \frac{|x \cap \hat{x}|}{|x \cup \hat{x}|} = \frac{|x \cap \hat{x}|}{|x| + |\hat{x}| - |x \cap \hat{x}|}
    \end{equation}

    The lower Jaccard similarity indicates that the model studied in the paper can effectively protect user privacy when generating data.

    \item \emph{Privacy budget and sensitivity.} In paper \cite{gao2024novel}, security is mainly guaranteed by differential privacy. Here is the Laplace mechanism formula for differential privacy.

    \begin{equation}
    M(D) = f(D) + \text{Lap} (\Delta f / \epsilon)
    \end{equation}

    where $\text{Lap} (\Delta f / \epsilon)$ is the noise sampled from the Laplace distribution, $\Delta f$ is the sensitivity of the query function, and $\epsilon$ is the privacy budget that reflects the level of privacy protection. The privacy budget is the core indicator of differential privacy and is used to measure the degree of privacy protection. The smaller $\epsilon$ is, the stronger the privacy protection is. The greater the sensitivity, the more sensitive the query results are to data changes, and more noise needs to be added to protect privacy. But the more noise is added, which may affect the availability of data and the accuracy of recommendations.

     \item \emph{T-HR@k and T-NDCG@k.} To evaluate the success rate of the attack, \cite{zhang2024lorec} employs metrics based on the top-$k$ performance of the target item, namely T-HR@k and T-NDCG@k. The specific formula for T-HR@k is as follows:

     \begin{equation}
     \text{T-HR}@k = \frac{1}{|\Gamma|}\sum_{tar\in\Gamma}\frac{\sum_u\in u_n-u_{n,tar}I(tar\in L_{u_{1:k})}}{|u_n-u_{n,tar}|}.
     \end{equation}
     
     Here, $\Gamma$ represents the set of target items, $U_{n,tar}$ is the set of genuine users who interacted with target item $tar$, $L_{u_{1:k}}$ is the top-$k$ recommendation list for user $u$, and $I(·)$ is an indicator function that returns 1 if the condition is true. T-NDCG@k is a target item-specific version of NDCG@k.

     \item \emph{$RC_{HR}@k$}. In \cite{zhang2024lorec}, $RC_{HR}@k$ is defined as a measure of robustness, which captures the variation in recommendation performance before and after an attack. It is calculated as:
     \begin{equation}
         \text{RC}_{\text{HR}}@k=1-\frac{|\text{HR}@k-\text{HR}@k_{clean}|}{\text{HR}@k_{clean}},
     \end{equation}
    where $\text{HR}@k_{clean}$ denotes the top-$k$ hit rate evaluated on the unattacked dataset. $\text{RC}_{\text{NDCG}}@k$ is computed in a similar manner, serving as the NDCG@k metric under clean data conditions.

    \item \emph{Effectiveness and Stealthiness}. Paper \cite{zhang2024stealthy} evaluates attacks with two key metrics: effectiveness and stealthiness. Effectiveness measures the extent to which an attack can promote a designated target item, typically quantified by item exposure or purchase propensity. Stealthiness, on the other hand, assesses how inconspicuous the attack is to both users and the platform. This is evaluated by examining whether overall recommendation performance remains largely unaffected, or by quantifying text quality and analyzing statistical patterns in the perturbation terms.
    
\end{itemize}

In the security evaluation of LLMs, Gao \emph{et al.} \cite{gao2024brief} and Chua \emph{et al.} \cite{chua2024ai} propose a variety of indicators specifically for measuring the security of model-generated content. The indicators are usually related to issues such as toxicity, bias, privacy leakage, and hallucination of model-generated content.

Different from conventional intuitive judgments, paper \cite{zhang2024misinformation} indirectly judges security by analyzing the spread and polarization of misinformation. The main focus is on the control of the spread of misinformation, the degree of extremism of group opinions, and the effect of algorithmic intervention.

\begin{itemize}
    \item \emph{Polarization.} It calculates the standard deviation of group opinions, reflecting the degree of divergence of group opinions. The higher the polarization, the more divided the group opinions are, which may lead to social instability.

    \item \emph{Radicalization.} It reflects the degree of extremism of group opinions. The higher the radicalization, the more extreme the group opinions are, which may lead to extreme behavior.

    \item \emph{Misinformation Spread.} It reflects the degree of spread of misinformation in social networks, which can be measured by calculating the length of the propagation path and the propagation speed of misinformation in social networks.

    \item \emph{Opinion Reversal Rate.} It reflects the reversal effect of algorithmic intervention on group opinions. The higher the reversal rate, the more effective the algorithmic intervention is, which can reduce the impact of misinformation.
\end{itemize}

\subsection{Other Objectives}
Some research objectives in MORS are relatively niche, with limited academic literature available. As such, this section will provide a comprehensive discussion of these objectives.

\begin{itemize}
%\item \textbf{Robustness.} \textbf{(1)Definition:} Robustness refers to the system's ability to maintain reliable and accurate performance under various challenging conditions, such as noisy data, adversarial attacks, or unexpected changes in user behavior or item availability. A robust recommendation system is resilient to these disturbances and continues to provide high-quality recommendations without significant degradation in performance. \textbf{(2)Methods:} To address a significant challenge called prompt sensitivity, which refers to that it is highly susceptible to the influence of prompt words, \cite{li2024ganprompt} proposes GANPrompt. By integrating GANs generation techniques to produce diverse prompt, this paper enhances the model’s adaptability and stability to unseen prompts, improving robustness of recommendation systems in complex and dynamic environments. \textbf{(3)Metrics:} To verify the robustness of the module, the paper \cite{li2024ganprompt} designed ablation experiments to compare the model's performance with and without the GAN module.
 \item \emph{Controllability.} (1) \textbf{Definition:} Controllability refers to the ability to adjust, regulate, or influence the behavior and outputs of the system according to specific requirements or constraints. This is essential to ensure that the recommendations align with desired goals, such as business objectives, ethical considerations, or user preferences, while maintaining transparency and accountability. (2) \textbf{Methods:} Paper \cite{chen2024dlcrec} enables precise control over diversity in LLM-based recommendations, with controlled genre numbers propagated through each stage to shape the final recommendation lists. In three sub-tasks, prompts of LLM will be modified to meet varying diversity requirements, where users can specify how many or exactly which genres the lists should contain. (3) \textbf{Metrics:} Paper \cite{chen2024dlcrec} uses $\textit{MAE}\_Cov@k$ to evaluate the model’s ability to meet the diversity requirements, which calculates the mean average error between the actual and desired genre coverage.

 \item \emph{Novelty.} (1) \textbf{Definition:} Novel items, by definition, do not have any user feedback data. A novel recommendation introduces users to items they are unlikely to have discovered on their own, thereby expanding their exposure to new content. Note that encouraging novelty in top-$k$ items is different from addressing the cold-start problem \cite{sanner2023large,wu2024could}. Novel items are defined $wrt.$ a quer,y whereas cold-start items are defined globally for a system. In many cases, a novel item $wrt.$ a query may be a popular item globally. (2) \textbf{Methods:} Past work uses approximations of novelty, such as an item being less popular \cite{shi2023relieving} or belonging to less popular categories \cite{xie2021hierarchical}, to enhance novelty, since truly novel items, those that have not been shown by the current system, would not have any user feedback data for training. In the work \cite{sharma2024optimizing}, they use the semantic capabilities of LLMs to propose a general, scalable method to estimate the relevance of novel items, providing feedback on novelty items to reinforcement learning. As a result, it can directly optimize for the novelty objective without sacrificing relevance. (3) \textbf{Metrics:} Novelty is often evaluated by measuring how different the recommended items are from the user's historical interactions. In \cite{sharma2024optimizing}, given a query $x$, $\textit{Novelty}@ k (x, \phi, \psi, L)$ is defined as the number of items in top-$k$ predictions of a candidate model $\phi$ that do not exist in top-$L$ predictions of the base model $\psi$.

 \item \emph{Efficiency.} (1) \textbf{Definition:} In the research on recommendation systems based on generative technologies, LLMs4Rec has emerged as a current focal point. However, due to the substantial computational costs, large storage requirements, and significant inference latency associated with LLMs during both training and inference, efficiency has gradually become a key optimization objective in LLM4Rec research. (2) \textbf{Methods:} In \cite{wu2024survey2}, the authors conducted a systematic survey of studies aimed at improving the efficiency of LLM4Rec, identifying three primary levels of optimization: the data level, which seeks to reduce processing costs by minimizing the data scale; the model level, which focuses on designing lighter and more efficient architectures to lower computational and storage overhead; and the strategy level, which emphasizes parameter-efficient fine-tuning methods tailored for recommendation tasks. As this work provides a comprehensive overview and summary of relevant approaches, this survey does not reiterate those details. Readers interested in further information are referred to the aforementioned study.
 \end{itemize}

\section{Datasets} \label{sec:exp}

\begin{table*}[h]
    \centering
    \renewcommand{\arraystretch}{1.5} % 调整行高
    \setlength{\tabcolsep}{0pt} % 调整列间距
    \caption{Overview of datasets}
    \begin{tabular}{ccc}
        \hline
        \textbf{Dataset}  & \textbf{References} & \textbf{Link}\\ \hline
        Anime  & \cite{wu2019pd,han2024controlling,carraro2024enhancing}  & https://www.kaggle.com/datasets/dbdmobile/myanimelist-dataset\\
        Goodreads & \cite{carraro2024enhancing} & https://www.kaggle.com/datasets/jealousleopard/goodreadsbooks\\
        Movielens & \cite{wu2019pd,chen2024dlcrec,han2024controlling,izawa2025diversifying,xu2023gs,borges2019enhancing,lilienthal2024synthetic,vcutura2024fairness,lilienthal2024multi,ning2024cheatagent,liu2025novel} & https://grouplens.org/datasets/movielens/ \\
        Steam & \cite{chen2024dlcrec} & https://www.kaggle.com/datasets/tamber/steam-video-games \\
        Amazon & \cite{izawa2025diversifying,xu2023gs,zhang2024gan,lilienthal2024synthetic,lilienthal2024multi,zhang2024lorec,zhang2024stealthy,deldjoo2021survey} & https://huggingface.co/datasets/McAuley-Lab/Amazon-Reviews-2023\\
        MIND & \cite{zhang2024lorec} & https://msnews.github.io/ \\
        Adressa & \cite{hassan2024gczrec} & https://reclab.idi.ntnu.no/dataset/ \\
        Book-Crossing & \cite{xu2023gs} & https://grouplens.org/datasets/book-crossing/ \\
        Serendipity-2018  & \cite{tokutake2024can} & https://grouplens.org/datasets/serendipity-2018/ \\
        SerenLens & \cite{fu2024art} & https://github.com/zhefu2/SerenLens/ \\
        Yelp  & \cite{xu2023gs,deldjoo2021survey} & https://www.kaggle.com/datasets/yelp-dataset/yelp-dataset\\
        German Credit   & \cite{kliachkin2024fairness1} & https://archive.ics.uci.edu/dataset/144/statlog+german+credit+data\\
        Netflix &   \cite{borges2019enhancing}  &   https://www.kaggle.com/datasets/netflix-inc/netflix-prize-data \\
        MSD    &   \cite{borges2019enhancing}   &   http://millionsongdataset.com/tasteprofile/ \\
        Reddit  & \cite{zhang2024misinformation} & https://snap.stanford.edu/graphsage/\#datasets \\
        RealToxicityPrompts & \cite{gao2024brief} & https://toxicdegeneration.allenai.org/ \\
        % Pinterest & \cite{deldjoo2021survey} & https://sites.google.com/site/xueatalphabeta/academic-projects\\
        TruthfulQA & \cite{gao2024brief} & https://github.com/sylinrl/TruthfulQA\\
        FACTOR & \cite{gao2024brief} & https://github.com/AI21Labs/factor \\
        HaDeS & \cite{gao2024brief} & https://github.com/microsoft/HaDes \\
        HalluQA & \cite{gao2024brief} & https://github.com/OpenMOSS/HalluQA \\
        HaluEval & \cite{gao2024brief} & https://github.com/RUCAIBox/HaluEval \\
        UHGEval & \cite{gao2024brief} & https://github.com/IAAR-Shanghai/UHGEval \\\hline
    \end{tabular}
    % \vspace{6pt}
\end{table*}

The datasets used in recommendation system research exhibit distinct characteristics depending on the specific objective. This section will categorize commonly used datasets across four dimensions: general recommendation, serendipity recommendation, fairness recommendation, and security recommendation. The relevant datasets, corresponding studies that utilize these datasets and database links are summarized in TABLE 2.

\subsection{General Recommendation Datasets}
General recommendation datasets refer to datasets that are widely applicable for evaluating various objectives in recommendation systems. Widely used datasets include MovieLens, Amazon and Yelp $etc$. These datasets contain rich user-item interactions, user demographic information, and item metadata, making them fundamental benchmarks for assessing recommendation algorithms. (1) \textbf{MovieLens.} It provides movie-related data, including movie titles, genres, explicit user ratings, timestamps, and user attributes such as age and gender. It supports research on collaborative filtering and temporal recommendation. The two commonly used subsets are MovieLens 1M and MovieLens 10M. MovieLens 1M dataset contains 1,000,209 anonymous ratings of approximately 3,900 movies made by 6,040 MovieLens users who joined MovieLens in 2000 while MovieLens 10M contains 10000054 ratings and 95580 tags applied to 10681 movies by 71567 users of the online movie recommendation service MovieLens. (2) \textbf{Amazon Review Dataset (hereafter Amazon)} The latest version is the 2023 edition. This dataset primarily collects user reviews, including ratings, text reviews and votes, as well as item metadata such as product descriptions, prices and images. It comprises multiple sub-datasets covering various commercial domains, including books, electronics and apparel. The choice of a specific sub-dataset depends on the research needs. (3) \textbf{The Yelp Open Dataset (hereafter Yelp).} It is a subset of business, review and user data from Yelp, a well-known business review platform. It comprises approximately 160,000 businesses, 8.63 million reviews and 200,000 images from eight major metropolitan areas. (4) \textbf{MIcrosoft News Dataset (MIND).} It is a large-scale dataset for news recommendation research. It contains about 160k English news articles and more than 15 million impression logs generated by 1 million users. 
%This dataset contains rich textual content, including the title, summary, body, category, and entities of each news article and also includes extensive implicit interaction data, such as each exposure log recording the user's prior click events, non-click events, and historical news click behavior.%Yelp,Amazon没找到对应的
(5) \textbf{Adressa.} It is another news dataset that includes news articles in Norwegian. It comes in two versions, the large 20M dataset of 10 weeks’ traffic on Adresseavisen’s news portal, and the small 2M dataset of only one week’s traffic \cite{gulla2017adressa}. (6) \textbf{Book-Crossing.} It comprises 278,858 anonymized users with demographic attributes, 271,379 books, and 1,149,780 user-book interactions, including explicit ratings and implicit feedback. (7) \textbf{Anime.} This dataset contains comprehensive details of 24,905 anime entries. (8) \textbf{Steam.} This is one of the world's most popular PC gaming platforms, featuring over 6,000 games and millions of players. This dataset captures user behavior, including columns such as user ID, game title, behavior type, value and some irrelevant data. The behavior types include ''purchase" and "play". When the behavior type is "purchase", the value is always 1, whereas for "play", the value represents the number of hours the user has spent playing the game. (9) \textbf{Goodreads.} This dataset is collected from the Goodreads API and includes approximately 45.6K books. It contains information such as book titles, authors, user ratings, languages and the number of reviews, making it a comprehensive resource for book-related data.

% \textcolor{blue}{(1) The names of the datasets do not correspond to those in Table 2. (I have already revised, please check if ok) (2) Some datasets mentioned here are not included in Table 2, such as "Adressa".}

\subsection{Datasets for Recommendation Serendipity }
Since serendipity is inherently subjective, obtaining genuinely serendipitous item data is crucial for evaluating serendipity in recommendation systems. Currently, two primary datasets have been specifically collected for this purpose, which are Serendipity 2018 and SerenLens. (1) \textbf{Serendipity-2018.} This dataset is derived from MovieLens and was constructed by conducting experiments in which participants were asked to identify movies they perceived as serendipitous. It is intended for research on serendipity in recommendation systems, including the analysis of serendipitous movies and the offline evaluation of serendipity-oriented recommendation algorithms. The interaction data in this dataset includes 104,661 users, 49,151 items and 9,997,850 ratings. (2) \textbf{SerenLens.} It is a collected large ground truth dataset on people’s serendipity experiences which is derived from Amazon Reviews Data. This multimodal dataset spans two domains, which are Books and Movies \& TV, with annotated reviews. The Books domain contains 265,037 reviews from 2,346 users across 113,876 unique book titles, while the Movies \& TV domain comprises 74,967 reviews contributed by 619 users evaluating 23,950 distinct movies/TV shows.

\subsection{Datasets for Recommendation Fairness }
As for fairness, the most commonly used datasets contain human characteristics, especially sensitive attributes such as gender, age, race, $etc$. (1) \textbf{German Credit.} This dataset is sourced from the UCI Machine Learning Repository and contains 1,000 samples. The dataset includes several attributes, with "Credit Amount" being the key basis for generating credit scores. In terms of protected attributes, the dataset combines Sex and Age into a single protected attribute (Sex-Age), which consists of four categories, such as " $<$ 35 years - Female" and "$\geq$ 35 years - Male". Additionally, housing status (Housing) is treated as an unknown protected attribute with three categories: "Own Housing", "Rented", and "Free Housing". In the experimental setup, Sex-Age is used as the known protected attribute to generate fair rankings, while Housing is used as the unknown protected attribute to assess the robustness of the algorithm \cite{kliachkin2024fairness1}. (2) \textbf{Netflix.} It contains movie ratings from 1998 to 2005. The original data contains more than 100 million rating records, involving 480,000 users and 17,000 movies. The data is also converted to binary interaction form, and users who interacted with less than 5 movies are removed. (3) \textbf{Million Song Dataset (MSD).} It is a large-scale music listening dataset that records users' song playing behavior. Users who have listened to less than 200 songs and songs that have been listened to by less than 20 users are excluded, providing ideal conditions for verifying the performance of the model in extremely sparse scenarios.

\subsection{Datasets for Recommendation Security}
In terms of security, multiple public datasets are used to evaluate the performance of the model in terms of privacy protection, recommendation systems and content generation. These datasets cover a variety of types, such as user ratings, social networks, and content tagging, and can fully reflect the capabilities of the model in different scenarios. (1) \textbf{Reddit.} This dataset is based on real interaction data from the Reddit social platform, including all submissions to the subreddit “r/news” from 2016 to 2020. The research team extracted discussion data on the topic of "TECH" in 2020 from the "r/news" sub-forum, and through rigorous screening and processing, ultimately formed a social network dataset containing 4585 user nodes and 6491 interaction edges to verify the effectiveness of the model in a real networks \cite{zhang2024misinformation}. (2) \textbf{RealToxicityPrompts.} This dataset is a large-scale dataset that focuses on detecting the tendency of language models to generate toxic content. Paper \cite{gao2024brief} states that the dataset contains web texts collected from untrustworthy websites with 4.3\% toxic content. By providing these potentially toxic prompts, researchers can evaluate the likelihood of the model generating insulting, hateful, or offensive language. (3) \textbf{TruthfulQA.} It is a benchmark dataset specifically designed to evaluate the authenticity of content generated by LLMs. By designing a series of questions that require factual answers, it tests whether the model will generate hallucinations that contradict objective facts. Because it pays special attention to the tendency of models to imitate common human wrong answers, such as generating seemingly reasonable but actually wrong answers to questions involving common sense, scientific facts, or historical events, this dataset has been widely used to detect the factual accuracy defects of models in open domain question answering tasks \cite{gao2024brief}. (4) \textbf{FACTOR.} This dataset detects the tendency of models to produce "similar but inaccurate" statements by comparing statements generated by the model with accurate information in the real corpus. It is an innovative automated evaluation framework whose core function is to transform the selected factual corpus into a quantifiable benchmark that can systematically measure the model's ability to distinguish subtle factual differences \cite{gao2024brief}. In addition, datasets such as \textbf{HaDes}, \textbf{HalluQA}, \textbf{HaluEval} and \textbf{UHGEval} are also used to evaluate the model’s performance in hallucination \cite{gao2024brief}.

\section{CHALLENGES AND OPPORTUNITIES} 
This paper has outlined several avenues for integrating generative models with various objectives in recommendation systems. However, challenges remain in several aspects. Here, we present the challenges and future directions.

(1) \textbf{Inconsistency in definitions and evaluation criteria:} Existing studies often confuse concepts such as diversity, novelty and serendipity, lacking clear definitions. Different studies have different understandings of fairness, and the definition of serendipity is highly subjective, lacking a unified evaluation framework. (2) \textbf{Data limitations:} Especially for goals such as serendipity and fairness, there is a lack of high-quality, standardized datasets. (3) \textbf{Challenges of generative technology:} Current research mainly focuses on LLMs and GANs, and the potential of other generative technologies such as diffusion models and VAEs has not been fully explored. At the same time, recommendation systems based on GAN or LLM may face problems such as training instability, data bias, transparency, and interpretability. Specially, the high latency of LLM reasoning makes it difficult to meet real-time recommendation requirements, affecting the feasibility of industrial deployment, and LLMs may have a gap in their understanding of concepts such as diversity and real user needs. (4) \textbf{Complexity of multi-objective optimization:} Existing research mainly focuses on the balance between accuracy and a single objective, while how to optimize multiple objectives simultaneously is still an open question. In the process of synergistic improvements, there may be conflicts between different goals (such as privacy protection may reduce the accuracy of recommendations), and more efficient multi-objective joint optimization methods need to be explored. (5) \textbf{Explainability and controllability:} At present, how to provide understandable explanations for generative recommendations is still a major challenge. The controllability mechanism is insufficient, that is, there is a lack of effective methods for users or system managers to regulate the recommended content generated by LLM. (6) \textbf{Security and privacy:} Privacy protection methods may reduce the quality of recommendations, and more efficient adaptive algorithms need to be designed. Recommendation systems face new types of attacks, but the research on defense mechanisms is still in its early stages. (7) \textbf{Ethics and user experience:} Generative recommendations may imply bias or discrimination, and an ethical review mechanism needs to be established. Improving system transparency and explainability to enhance user trust is the key to practical applications.

Based on the above challenges, we propose the following possible future development directions.

\begin{itemize}
    \item Establish standardized definitions and evaluation frameworks, especially for highly subjective goals such as serendipity.
    \item Develop multi-objective optimization methods that take into account more objectives.
    \item Explore new technologies such as diffusion models and VAEs, or combine traditional collaborative filtering methods to improve recommendation performance.
    \item Optimize LLM reasoning efficiency and reduce latency to meet the needs of industrial scenarios.
    \item Enhance explainability and controllability, design user-intervenible recommendation generation technologies, and provide intuitive explanations.
    \item Enhance privacy and security, develop adaptive privacy protection algorithms, and establish a defense framework for adversarial attacks.
    \item Carry out cross-domain cooperation and promote data sharing and technology standardization through open source benchmarks and cross-industry collaboration.
\end{itemize}
\section{Conclusion}
This paper provides an extensive survey of generative AI technologies applied to MORS. As the applications of generative methods for multi-objective optimization in recommendation systems remains relatively underexplored, this study focused on the usage of generative techniques across several key objectives: diversity, serendipity, fairness and safety. By examining current research in each of these areas, the paper illustrates the potential of generative models in addressing the inherent challenges and enhancing the performance of recommendation systems. For each objective, this survey identifies commonly employed evaluation methods and datasets, offering a comprehensive overview of the current state of research.

While the applications of generative technologies in MORS show significant promise, it is clear that further research is needed to unlock their full potential. 
Addressing the outlined challenges and advancing optimization methodologies will be key to driving progress in this domain.

\ifCLASSOPTIONcaptionsoff
  \newpage
\fi

% % trigger a \newpage just before the given reference
% % number - used to balance the columns on the last page
% % adjust value as needed - may need to be readjusted if
% % the document is modified later
% %\IEEEtriggeratref{8}
% % The "triggered" command can be changed if desired:
% %\IEEEtriggercmd{\enlargethispage{-5in}}

% % references section

% % can use a bibliography generated by BibTeX as a .bbl file
% % BibTeX documentation can be easily obtained at:
% % http://mirror.ctan.org/biblio/bibtex/contrib/doc/
% % The IEEEtran BibTeX style support page is at:
% % http://www.michaelshell.org/tex/ieeetran/bibtex/
% %\bibliographystyle{IEEEtran}
% % argument is your BibTeX string definitions and bibliography database(s)
% %\bibliography{IEEEabrv,../bib/paper}
% %
% % <OR> manually copy in the resultant .bbl file
% % set second argument of \begin to the number of references
% % (used to reserve space for the reference number labels box)
% \begin{thebibliography}{1}

% \bibitem{IEEEhowto:kopka}
% H.~Kopka and P.~W. Daly, \emph{A Guide to \LaTeX}, 3rd~ed.\hskip 1em plus
%   0.5em minus 0.4em\relax Harlow, England: Addison-Wesley, 1999.

% \end{thebibliography}

% \bibliographystyle{IEEEtran}
\bibliography{main}

\begin{thebibliography}{100}

\bibitem{dong2024musechat}
Zhikang Dong, Xiulong Liu, Bin Chen, Pawel Polak, and Peng Zhang.
\newblock Musechat: A conversational music recommendation system for videos.
\newblock In {\em Proceedings of the IEEE/CVF Conference on Computer Vision and Pattern Recognition}, pages 12775--12785, 2024.

\bibitem{li2024review}
Xinmiao Li, Shuang Feng, and Xin Zhang.
\newblock A review of methods using large language models in news recommendation systems.
\newblock In {\em 2024 IEEE/ACIS 27th International Conference on Software Engineering, Artificial Intelligence, Networking and Parallel/Distributed Computing (SNPD)}, pages 81--86. IEEE, 2024.

\bibitem{patil2023survey}
Atharva Patil, Divyansh Suwalka, Aryan Kumar, Gaurav Rai, and Jayita Saha.
\newblock A survey on artificial intelligence (ai) based job recommendation systems.
\newblock In {\em 2023 International Conference on Sustainable Computing and Data Communication Systems (ICSCDS)}, pages 730--737. IEEE, 2023.

\bibitem{schafer2007collaborative}
J~Ben Schafer, Dan Frankowski, Jon Herlocker, and Shilad Sen.
\newblock Collaborative filtering recommender systems.
\newblock In {\em The adaptive web: methods and strategies of web personalization}, pages 291--324. Springer, 2007.

\bibitem{herlocker2004evaluating}
Jonathan~L Herlocker, Joseph~A Konstan, Loren~G Terveen, and John~T Riedl.
\newblock Evaluating collaborative filtering recommender systems.
\newblock {\em ACM Transactions on Information Systems (TOIS)}, 22(1):5--53, 2004.

\bibitem{papadakis2022collaborative}
Harris Papadakis, Antonis Papagrigoriou, Costas Panagiotakis, Eleftherios Kosmas, and Paraskevi Fragopoulou.
\newblock Collaborative filtering recommender systems taxonomy.
\newblock {\em Knowledge and Information Systems}, 64(1):35--74, 2022.

\bibitem{koren2021advances}
Yehuda Koren, Steffen Rendle, and Robert Bell.
\newblock Advances in collaborative filtering.
\newblock {\em Recommender systems handbook}, pages 91--142, 2021.

\bibitem{lops2011content}
Pasquale Lops, Marco De~Gemmis, and Giovanni Semeraro.
\newblock Content-based recommender systems: State of the art and trends.
\newblock {\em Recommender systems handbook}, pages 73--105, 2011.

\bibitem{javed2021review}
Umair Javed, Kamran Shaukat, Ibrahim~A Hameed, Farhat Iqbal, Talha~Mahboob Alam, and Suhuai Luo.
\newblock A review of content-based and context-based recommendation systems.
\newblock {\em International Journal of Emerging Technologies in Learning (iJET)}, 16(3):274--306, 2021.

\bibitem{dong2017hybrid}
Xin Dong, Lei Yu, Zhonghuo Wu, Yuxia Sun, Lingfeng Yuan, and Fangxi Zhang.
\newblock A hybrid collaborative filtering model with deep structure for recommender systems.
\newblock In {\em Proceedings of the AAAI Conference on artificial intelligence}, volume~31, 2017.

\bibitem{fan2020graph}
Wenqi Fan, Yao Ma, Qing Li, Jianping Wang, Guoyong Cai, Jiliang Tang, and Dawei Yin.
\newblock A graph neural network framework for social recommendations.
\newblock {\em IEEE Transactions on Knowledge and Data Engineering}, 34(5):2033--2047, 2020.

\bibitem{zhang2019deep}
Shuai Zhang, Lina Yao, Aixin Sun, and Yi~Tay.
\newblock Deep learning based recommender system: A survey and new perspectives.
\newblock {\em ACM computing surveys (CSUR)}, 52(1):1--38, 2019.

\bibitem{naumov2019deep}
Maxim Naumov, Dheevatsa Mudigere, Hao-Jun~Michael Shi, Jianyu Huang, Narayanan Sundaraman, Jongsoo Park, Xiaodong Wang, Udit Gupta, Carole-Jean Wu, Alisson~G Azzolini, et~al.
\newblock Deep learning recommendation model for personalization and recommendation systems.
\newblock {\em arXiv preprint arXiv:1906.00091}, 2019.

\bibitem{wang2023generative}
Wenjie Wang, Xinyu Lin, Fuli Feng, Xiangnan He, and Tat-Seng Chua.
\newblock Generative recommendation: Towards next-generation recommender paradigm.
\newblock {\em arXiv preprint arXiv:2304.03516}, 2023.

\bibitem{goodfellow2014generative}
Ian Goodfellow, Jean Pouget-Abadie, Mehdi Mirza, Bing Xu, David Warde-Farley, Sherjil Ozair, Aaron Courville, and Yoshua Bengio.
\newblock Generative adversarial nets.
\newblock {\em Advances in neural information processing systems}, 27, 2014.

\bibitem{liang2024survey}
Shangsong Liang, Zhou Pan, wei liu, Jian Yin, and Maarten de~Rijke.
\newblock A survey on variational autoencoders in recommender systems.
\newblock {\em ACM Computing Surveys}, 2024.

\bibitem{wang2023diffusion}
Wenjie Wang, Yiyan Xu, Fuli Feng, Xinyu Lin, Xiangnan He, and Tat-Seng Chua.
\newblock Diffusion recommender model.
\newblock In {\em Proceedings of the 46th International ACM SIGIR Conference on Research and Development in Information Retrieval}, pages 832--841, 2023.

\bibitem{wu2024survey}
Likang Wu, Zhi Zheng, Zhaopeng Qiu, Hao Wang, Hongchao Gu, Tingjia Shen, Chuan Qin, Chen Zhu, Hengshu Zhu, Qi~Liu, et~al.
\newblock A survey on large language models for recommendation.
\newblock {\em World Wide Web}, 27(5):60, 2024.

\bibitem{nguyen2014exploring}
Tien~T Nguyen, Pik-Mai Hui, F~Maxwell Harper, Loren Terveen, and Joseph~A Konstan.
\newblock Exploring the filter bubble: the effect of using recommender systems on content diversity.
\newblock In {\em Proceedings of the 23rd international conference on World wide web}, pages 677--686, 2014.

\bibitem{jannach2022multi}
Dietmar Jannach.
\newblock Multi-objective recommender systems: Survey and challenges.
\newblock {\em arXiv preprint arXiv:2210.10309}, 2022.

\bibitem{Neurocomputing_zheng2022survey}
Yong Zheng and David~Xuejun Wang.
\newblock A survey of recommender systems with multi-objective optimization.
\newblock {\em Neurocomputing}, 474:141--153, 2022.

\bibitem{FBD_jannach2023survey}
Dietmar Jannach and Himan Abdollahpouri.
\newblock A survey on multi-objective recommender systems.
\newblock {\em Frontiers in big Data}, 6:1157899, 2023.

\bibitem{kaminskas2016diversity}
Marius Kaminskas and Derek Bridge.
\newblock Diversity, serendipity, novelty, and coverage: a survey and empirical analysis of beyond-accuracy objectives in recommender systems.
\newblock {\em ACM Transactions on Interactive Intelligent Systems (TiiS)}, 7(1):1--42, 2016.

\bibitem{roy2022systematic}
Deepjyoti Roy and Mala Dutta.
\newblock A systematic review and research perspective on recommender systems.
\newblock {\em Journal of Big Data}, 9(1):59, 2022.

\bibitem{isinkaye2015recommendation}
Folasade~Olubusola Isinkaye, Yetunde~O Folajimi, and Bolande~Adefowoke Ojokoh.
\newblock Recommendation systems: Principles, methods and evaluation.
\newblock {\em Egyptian informatics journal}, 16(3):261--273, 2015.

\bibitem{kumar2018recommendation}
Pushpendra Kumar and Ramjeevan~Singh Thakur.
\newblock Recommendation system techniques and related issues: a survey.
\newblock {\em International Journal of Information Technology}, 10:495--501, 2018.

\bibitem{wu2022graph}
Shiwen Wu, Fei Sun, Wentao Zhang, Xu~Xie, and Bin Cui.
\newblock Graph neural networks in recommender systems: a survey.
\newblock {\em ACM Computing Surveys}, 55(5):1--37, 2022.

\bibitem{gao2023survey}
Chen Gao, Yu~Zheng, Nian Li, Yinfeng Li, Yingrong Qin, Jinghua Piao, Yuhan Quan, Jianxin Chang, Depeng Jin, Xiangnan He, et~al.
\newblock A survey of graph neural networks for recommender systems: Challenges, methods, and directions.
\newblock {\em ACM Transactions on Recommender Systems}, 1(1):1--51, 2023.

\bibitem{zhang2021artificial}
Qian Zhang, Jie Lu, and Yaochu Jin.
\newblock Artificial intelligence in recommender systems.
\newblock {\em Complex \& Intelligent Systems}, 7(1):439--457, 2021.

\bibitem{masciari2024systematic}
Elio Masciari, Areeba Umair, and Muhammad~Habib Ullah.
\newblock A systematic literature review on ai based recommendation systems and their ethical considerations.
\newblock {\em IEEE Access}, 2024.

\bibitem{bharadhwaj2018recgan}
Homanga Bharadhwaj, Homin Park, and Brian~Y Lim.
\newblock Recgan: recurrent generative adversarial networks for recommendation systems.
\newblock In {\em Proceedings of the 12th ACM Conference on Recommender Systems}, pages 372--376, 2018.

\bibitem{liu2020deep}
Huafeng Liu, Liping Jing, Jingxuan Wen, Zhicheng Wu, Xiaoyi Sun, Jiaqi Wang, Lin Xiao, and Jian Yu.
\newblock Deep global and local generative model for recommendation.
\newblock In {\em Proceedings of the web conference 2020}, pages 551--561, 2020.

\bibitem{zheng2023generative}
Zhi Zheng, Zhaopeng Qiu, Xiao Hu, Likang Wu, Hengshu Zhu, and Hui Xiong.
\newblock Generative job recommendations with large language model.
\newblock {\em arXiv preprint arXiv:2307.02157}, 2023.

\bibitem{gao2024generative}
Shen Gao, Jiabao Fang, Quan Tu, Zhitao Yao, Zhumin Chen, Pengjie Ren, and Zhaochun Ren.
\newblock Generative news recommendation.
\newblock In {\em Proceedings of the ACM Web Conference 2024}, pages 3444--3453, 2024.

\bibitem{li2023gpt4rec}
Jinming Li, Wentao Zhang, Tian Wang, Guanglei Xiong, Alan Lu, and Gerard Medioni.
\newblock Gpt4rec: A generative framework for personalized recommendation and user interests interpretation.
\newblock {\em arXiv preprint arXiv:2304.03879}, 2023.

\bibitem{walker2022recommendation}
Joojo Walker, Ting Zhong, Fengli Zhang, Qiang Gao, and Fan Zhou.
\newblock Recommendation via collaborative diffusion generative model.
\newblock In {\em International Conference on Knowledge Science, Engineering and Management}, pages 593--605. Springer, 2022.

\bibitem{gao2021recommender}
Min Gao, Junwei Zhang, Junliang Yu, Jundong Li, Junhao Wen, and Qingyu Xiong.
\newblock Recommender systems based on generative adversarial networks: A problem-driven perspective.
\newblock {\em Information Sciences}, 546:1166--1185, 2021.

\bibitem{li2023large}
Lei Li, Yongfeng Zhang, Dugang Liu, and Li~Chen.
\newblock Large language models for generative recommendation: A survey and visionary discussions.
\newblock {\em arXiv preprint arXiv:2309.01157}, 2023.

\bibitem{zhao2023recommender}
Zihuai Zhao, Wenqi Fan, Jiatong Li, Yunqing Liu, Xiaowei Mei, Yiqi Wang, Zhen Wen, Fei Wang, Xiangyu Zhao, Jiliang Tang, et~al.
\newblock Recommender systems in the era of large language models (llms).
\newblock {\em arXiv preprint arXiv:2307.02046}, 2023.

\bibitem{deldjoo2024review}
Yashar Deldjoo, Zhankui He, Julian McAuley, Anton Korikov, Scott Sanner, Arnau Ramisa, Ren{\'e} Vidal, Maheswaran Sathiamoorthy, Atoosa Kasirzadeh, and Silvia Milano.
\newblock A review of modern recommender systems using generative models (gen-recsys).
\newblock In {\em Proceedings of the 30th ACM SIGKDD Conference on Knowledge Discovery and Data Mining}, pages 6448--6458, 2024.

\bibitem{li2024survey}
Yongqi Li, Xinyu Lin, Wenjie Wang, Fuli Feng, Liang Pang, Wenjie Li, Liqiang Nie, Xiangnan He, and Tat-Seng Chua.
\newblock A survey of generative search and recommendation in the era of large language models.
\newblock {\em arXiv preprint arXiv:2404.16924}, 2024.

\bibitem{ayemowa2024analysis}
Matthew~O Ayemowa, Roliana Ibrahim, and Muhammad~Murad Khan.
\newblock Analysis of recommender system using generative artificial intelligence: A systematic literature review.
\newblock {\em IEEE Access}, 2024.

\bibitem{wang2024enhanced}
Yidan Wang, Zhaochun Ren, Weiwei Sun, Jiyuan Yang, Zhixiang Liang, Xin Chen, Ruobing Xie, Su~Yan, Xu~Zhang, Pengjie Ren, et~al.
\newblock Enhanced generative recommendation via content and collaboration integration.
\newblock {\em arXiv e-prints}, pages arXiv--2403, 2024.

\bibitem{lin2024survey}
Jianghao Lin, Jiaqi Liu, Jiachen Zhu, Yunjia Xi, Chengkai Liu, Yangtian Zhang, Yong Yu, and Weinan Zhang.
\newblock A survey on diffusion models for recommender systems.
\newblock {\em arXiv preprint arXiv:2409.05033}, 2024.

\bibitem{vats2024exploring}
Arpita Vats, Vinija Jain, Rahul Raja, and Aman Chadha.
\newblock Exploring the impact of large language models on recommender systems: An extensive review.
\newblock {\em arXiv preprint arXiv:2402.18590}, 2024.

\bibitem{lin2025can}
Jianghao Lin, Xinyi Dai, Yunjia Xi, Weiwen Liu, Bo~Chen, Hao Zhang, Yong Liu, Chuhan Wu, Xiangyang Li, Chenxu Zhu, et~al.
\newblock How can recommender systems benefit from large language models: A survey.
\newblock {\em ACM Transactions on Information Systems}, 43(2):1--47, 2025.

\bibitem{xu2024large}
Derong Xu, Wei Chen, Wenjun Peng, Chao Zhang, Tong Xu, Xiangyu Zhao, Xian Wu, Yefeng Zheng, Yang Wang, and Enhong Chen.
\newblock Large language models for generative information extraction: A survey.
\newblock {\em Frontiers of Computer Science}, 18(6):186357, 2024.

\bibitem{alhijawi2022survey}
Bushra Alhijawi, Arafat Awajan, and Salam Fraihat.
\newblock Survey on the objectives of recommender systems: Measures, solutions, evaluation methodology, and new perspectives.
\newblock {\em ACM Computing Surveys}, 55(5):1--38, 2022.

\bibitem{kunaver2017diversity}
Matev{\v{z}} Kunaver and Toma{\v{z}} Po{\v{z}}rl.
\newblock Diversity in recommender systems--a survey.
\newblock {\em Knowledge-based systems}, 123:154--162, 2017.

\bibitem{mendoza2020evaluating}
Marcelo Mendoza and Nicol{\'a}s Torres.
\newblock Evaluating content novelty in recommender systems.
\newblock {\em Journal of Intelligent Information Systems}, 54:297--316, 2020.

\bibitem{kotkov2016survey}
Denis Kotkov, Shuaiqiang Wang, and Jari Veijalainen.
\newblock A survey of serendipity in recommender systems.
\newblock {\em Knowledge-Based Systems}, 111:180--192, 2016.

\bibitem{kotkov2023rethinking}
Denis Kotkov, Alan Medlar, and Dorota Glowacka.
\newblock Rethinking serendipity in recommender systems.
\newblock In {\em Proceedings of the 2023 Conference on Human Information Interaction and Retrieval}, pages 383--387, 2023.

\bibitem{himeur2022latest}
Yassine Himeur, Shahab~Saquib Sohail, Faycal Bensaali, Abbes Amira, and Mamoun Alazab.
\newblock Latest trends of security and privacy in recommender systems: a comprehensive review and future perspectives.
\newblock {\em Computers \& Security}, 118:102746, 2022.

\bibitem{wang2023survey}
Yifan Wang, Weizhi Ma, Min Zhang, Yiqun Liu, and Shaoping Ma.
\newblock A survey on the fairness of recommender systems.
\newblock {\em ACM Transactions on Information Systems}, 41(3):1--43, 2023.

\bibitem{chen2023bias}
Jiawei Chen, Hande Dong, Xiang Wang, Fuli Feng, Meng Wang, and Xiangnan He.
\newblock Bias and debias in recommender system: A survey and future directions.
\newblock {\em ACM Transactions on Information Systems}, 41(3):1--39, 2023.

\bibitem{zaizi2023comparative}
Fatima~Ezzahra Zaizi, Sara Qassimi, and Said Rakrak.
\newblock A comparative study of evolutionary algorithms for multi-objective recommender systems.
\newblock In {\em 2023 9th International Conference on Optimization and Applications (ICOA)}, pages 1--6. IEEE, 2023.

\bibitem{perera2023multiple}
Yomal Perera, Rukshan Karannagoda, Dion Weiman, and Subha Fernando.
\newblock Multiple objective optimization based dietary recommender system.
\newblock In {\em 2023 8th International Conference on Information Technology Research (ICITR)}, pages 1--6. IEEE, 2023.

\bibitem{zaizi2024towards}
Fatima~Ezzahra Zaizi, Sara Qassimi, and Said Rakrak.
\newblock Towards an efficient tourism recommendation system using multi-objective evolutionary optimization.
\newblock In {\em 2024 International Conference on Control, Automation and Diagnosis (ICCAD)}, pages 1--6. IEEE, 2024.

\bibitem{zhang2022community}
Lei Zhang, Huabin Zhang, Sibo Liu, Chao Wang, and Hongke Zhao.
\newblock A community division-based evolutionary algorithm for large-scale multi-objective recommendations.
\newblock {\em IEEE Transactions on Emerging Topics in Computational Intelligence}, 7(5):1470--1483, 2022.

\bibitem{zhang2021balancing}
Langlang Zhang, Anqi Pan, and Hongrui Shi.
\newblock Balancing diversity and accuracy of the recommendation system based on multi-objective optimization.
\newblock In {\em 2021 China Automation Congress (CAC)}, pages 542--547. IEEE, 2021.

\bibitem{zhou2023dynamic}
Wei Zhou, Yong Liu, Min Li, Yu~Wang, Zhiqi Shen, Liang Feng, and Zexuan Zhu.
\newblock Dynamic multi-objective optimization framework with interactive evolution for sequential recommendation.
\newblock {\em IEEE Transactions on Emerging Topics in Computational Intelligence}, 7(4):1228--1241, 2023.

\bibitem{zhou2024diversified}
Wei Zhou, Xiaolong Luo, Hongyue Bao, and Zexuan Zhu.
\newblock Diversified sequential recommendation via evolutionary multi-objective transfer optimization.
\newblock In {\em 2024 IEEE Conference on Artificial Intelligence (CAI)}, pages 456--457. IEEE, 2024.

\bibitem{zaizi2023multi}
Fatima~Ezzahra Zaizi, Sara Qassimi, and Said Rakrak.
\newblock Multi-objective optimization with recommender systems: A systematic review.
\newblock {\em Information Systems}, 117:102233, 2023.

\bibitem{liu2024large}
Wanyi Liu, Long Chen, and Zhenzhou Tang.
\newblock Large language model aided multi-objective evolutionary algorithm: a low-cost adaptive approach.
\newblock {\em arXiv preprint arXiv:2410.02301}, 2024.

\bibitem{brock2018large}
Andrew Brock, Jeff Donahue, and Karen Simonyan.
\newblock Large scale gan training for high fidelity natural image synthesis.
\newblock {\em arXiv preprint arXiv:1809.11096}, 2018.

\bibitem{zhan2019spatial}
Fangneng Zhan, Hongyuan Zhu, and Shijian Lu.
\newblock Spatial fusion gan for image synthesis.
\newblock In {\em Proceedings of the IEEE/CVF conference on computer vision and pattern recognition}, pages 3653--3662, 2019.

\bibitem{qin2020gan}
Zhiwei Qin, Zhao Liu, Ping Zhu, and Yongbo Xue.
\newblock A gan-based image synthesis method for skin lesion classification.
\newblock {\em Computer methods and programs in biomedicine}, 195:105568, 2020.

\bibitem{azadi2018multi}
Samaneh Azadi, Matthew Fisher, Vladimir~G Kim, Zhaowen Wang, Eli Shechtman, and Trevor Darrell.
\newblock Multi-content gan for few-shot font style transfer.
\newblock In {\em Proceedings of the IEEE conference on computer vision and pattern recognition}, pages 7564--7573, 2018.

\bibitem{xu2021drb}
Wenju Xu, Chengjiang Long, Ruisheng Wang, and Guanghui Wang.
\newblock Drb-gan: A dynamic resblock generative adversarial network for artistic style transfer.
\newblock In {\em Proceedings of the IEEE/CVF international conference on computer vision}, pages 6383--6392, 2021.

\bibitem{tran2021data}
Ngoc-Trung Tran, Viet-Hung Tran, Ngoc-Bao Nguyen, Trung-Kien Nguyen, and Ngai-Man Cheung.
\newblock On data augmentation for gan training.
\newblock {\em IEEE Transactions on Image Processing}, 30:1882--1897, 2021.

\bibitem{frid2018synthetic}
Maayan Frid-Adar, Eyal Klang, Michal Amitai, Jacob Goldberger, and Hayit Greenspan.
\newblock Synthetic data augmentation using gan for improved liver lesion classification.
\newblock In {\em 2018 IEEE 15th international symposium on biomedical imaging (ISBI 2018)}, pages 289--293. IEEE, 2018.

\bibitem{mariani2018bagan}
Giovanni Mariani, Florian Scheidegger, Roxana Istrate, Costas Bekas, and Cristiano Malossi.
\newblock Bagan: Data augmentation with balancing gan.
\newblock {\em arXiv preprint arXiv:1803.09655}, 2018.

\bibitem{lee2019collagan}
Dongwook Lee, Junyoung Kim, Won-Jin Moon, and Jong~Chul Ye.
\newblock Collagan: Collaborative gan for missing image data imputation.
\newblock In {\em Proceedings of the IEEE/CVF conference on computer vision and pattern recognition}, pages 2487--2496, 2019.

\bibitem{yoon2018gain}
Jinsung Yoon, James Jordon, and Mihaela Schaar.
\newblock Gain: Missing data imputation using generative adversarial nets.
\newblock In {\em International conference on machine learning}, pages 5689--5698. PMLR, 2018.

\bibitem{zhang2018metagan}
Ruixiang Zhang, Tong Che, Zoubin Ghahramani, Yoshua Bengio, and Yangqiu Song.
\newblock Metagan: An adversarial approach to few-shot learning.
\newblock {\em Advances in neural information processing systems}, 31, 2018.

\bibitem{wang2021siamese}
Qingxian Wang, Renjian Zhang, Kangkang Ma, Bo~Chen, Jiufang Chen, and Xiaoyu Shi.
\newblock Siamese generative adversarial predicting network for extremely sparse data in recommendation system.
\newblock In {\em 2021 IEEE Intl Conf on Parallel \& Distributed Processing with Applications, Big Data \& Cloud Computing, Sustainable Computing \& Communications, Social Computing \& Networking (ISPA/BDCloud/SocialCom/SustainCom)}, pages 1236--1241. IEEE, 2021.

\bibitem{woo2022conditional}
Sungpil Woo, Muhammad Zubair, Sunhwan Lim, Chan-Won Park, and Daeyoung Kim.
\newblock Conditional gan based collaborative filtering with data augmentation for cold-start user.
\newblock In {\em 2022 13th International Conference on Information and Communication Technology Convergence (ICTC)}, pages 1756--1761. IEEE, 2022.

\bibitem{chen2024collaborative}
JunShu Chen and GuangCong Liu.
\newblock Collaborative filtering algorithm based on generative adversarial networks.
\newblock In {\em 2024 IEEE 14th International Conference on Electronics Information and Emergency Communication (ICEIEC)}, pages 1--6. IEEE, 2024.

\bibitem{shafqat2022hybrid}
Wafa Shafqat and Yung-Cheol Byun.
\newblock A hybrid gan-based approach to solve imbalanced data problem in recommendation systems.
\newblock {\em IEEE access}, 10:11036--11047, 2022.

\bibitem{zhou2023vcgan}
Chung-Han Zhou and Yi-Ling Chen.
\newblock Vcgan: Variational collaborative generative adversarial network for recommendation systems.
\newblock In {\em ICC 2023-IEEE International Conference on Communications}, pages 6324--6330. IEEE, 2023.

\bibitem{ho2020denoising}
Jonathan Ho, Ajay Jain, and Pieter Abbeel.
\newblock Denoising diffusion probabilistic models.
\newblock {\em Advances in neural information processing systems}, 33:6840--6851, 2020.

\bibitem{wu2024diffusion}
Tengqing Wu.
\newblock A diffusion data enhancement retentive model for sequential recommendation.
\newblock In {\em 2024 7th International Conference on Computer Information Science and Application Technology (CISAT)}, pages 114--118. IEEE, 2024.

\bibitem{you2024context}
Di~You and Kyumin Lee.
\newblock Context-aware diffusion-based sequential recommendation.
\newblock In {\em 2024 IEEE International Conference on Big Data (BigData)}, pages 670--679. IEEE, 2024.

\bibitem{wu2019neural}
Le~Wu, Peijie Sun, Yanjie Fu, Richang Hong, Xiting Wang, and Meng Wang.
\newblock A neural influence diffusion model for social recommendation.
\newblock In {\em Proceedings of the 42nd international ACM SIGIR conference on research and development in information retrieval}, pages 235--244, 2019.

\bibitem{wu2020diffnet++}
Le~Wu, Junwei Li, Peijie Sun, Richang Hong, Yong Ge, and Meng Wang.
\newblock Diffnet++: A neural influence and interest diffusion network for social recommendation.
\newblock {\em IEEE Transactions on Knowledge and Data Engineering}, 34(10):4753--4766, 2020.

\bibitem{chen2024g}
Ruixin Chen, Jianping Fan, Meiqin Wu, Rui Cheng, and Jiawen Song.
\newblock G-diff: A graph-based decoding network for diffusion recommender model.
\newblock {\em IEEE Transactions on Neural Networks and Learning Systems}, 2024.

\bibitem{xu2024diffusion}
Yiyan Xu, Wenjie Wang, Fuli Feng, Yunshan Ma, Jizhi Zhang, and Xiangnan He.
\newblock Diffusion models for generative outfit recommendation.
\newblock In {\em Proceedings of the 47th international ACM SIGIR conference on research and development in information retrieval}, pages 1350--1359, 2024.

\bibitem{harte2023leveraging}
Jesse Harte, Wouter Zorgdrager, Panos Louridas, Asterios Katsifodimos, Dietmar Jannach, and Marios Fragkoulis.
\newblock Leveraging large language models for sequential recommendation.
\newblock In {\em Proceedings of the 17th ACM Conference on Recommender Systems}, pages 1096--1102, 2023.

\bibitem{friedman2023leveraging}
Luke Friedman, Sameer Ahuja, David Allen, Zhenning Tan, Hakim Sidahmed, Changbo Long, Jun Xie, Gabriel Schubiner, Ajay Patel, Harsh Lara, et~al.
\newblock Leveraging large language models in conversational recommender systems.
\newblock {\em arXiv preprint arXiv:2305.07961}, 2023.

\bibitem{chu2023leveraging}
Zhixuan Chu, Hongyan Hao, Xin Ouyang, Simeng Wang, Yan Wang, Yue Shen, Jinjie Gu, Qing Cui, Longfei Li, Siqiao Xue, et~al.
\newblock Leveraging large language models for pre-trained recommender systems.
\newblock {\em arXiv preprint arXiv:2308.10837}, 2023.

\bibitem{liu2023pre}
Peng Liu, Lemei Zhang, and Jon~Atle Gulla.
\newblock Pre-train, prompt, and recommendation: A comprehensive survey of language modeling paradigm adaptations in recommender systems.
\newblock {\em Transactions of the Association for Computational Linguistics}, 11:1553--1571, 2023.

\bibitem{wu2020ptum}
Chuhan Wu, Fangzhao Wu, Tao Qi, Jianxun Lian, Yongfeng Huang, and Xing Xie.
\newblock Ptum: Pre-training user model from unlabeled user behaviors via self-supervision.
\newblock {\em arXiv preprint arXiv:2010.01494}, 2020.

\bibitem{cui2022m6}
Zeyu Cui, Jianxin Ma, Chang Zhou, Jingren Zhou, and Hongxia Yang.
\newblock M6-rec: Generative pretrained language models are open-ended recommender systems.
\newblock {\em arXiv preprint arXiv:2205.08084}, 2022.

\bibitem{kim2021intent}
Hiun Kim, Jisu Jeong, Kyung-Min Kim, Dongjun Lee, Hyun~Dong Lee, Dongpil Seo, Jeeseung Han, Dong~Wook Park, Ji~Ae Heo, and Rak~Yeong Kim.
\newblock Intent-based product collections for e-commerce using pretrained language models.
\newblock In {\em 2021 International Conference on Data Mining Workshops (ICDMW)}, pages 228--237. IEEE, 2021.

\bibitem{mao2023unitrec}
Zhiming Mao, Huimin Wang, Yiming Du, and Kam-Fai Wong.
\newblock Unitrec: A unified text-to-text transformer and joint contrastive learning framework for text-based recommendation.
\newblock {\em arXiv preprint arXiv:2305.15756}, 2023.

\bibitem{bao2023tallrec}
Keqin Bao, Jizhi Zhang, Yang Zhang, Wenjie Wang, Fuli Feng, and Xiangnan He.
\newblock Tallrec: An effective and efficient tuning framework to align large language model with recommendation.
\newblock In {\em Proceedings of the 17th ACM Conference on Recommender Systems}, pages 1007--1014, 2023.

\bibitem{mann2020language}
Ben Mann, N~Ryder, M~Subbiah, J~Kaplan, P~Dhariwal, A~Neelakantan, P~Shyam, G~Sastry, A~Askell, S~Agarwal, et~al.
\newblock Language models are few-shot learners.
\newblock {\em arXiv preprint arXiv:2005.14165}, 1:3, 2020.

\bibitem{wei2022chain}
Jason Wei, Xuezhi Wang, Dale Schuurmans, Maarten Bosma, Fei Xia, Ed~Chi, Quoc~V Le, Denny Zhou, et~al.
\newblock Chain-of-thought prompting elicits reasoning in large language models.
\newblock {\em Advances in neural information processing systems}, 35:24824--24837, 2022.

\bibitem{liu2023chatgpt}
Junling Liu, Chao Liu, Peilin Zhou, Renjie Lv, Kang Zhou, and Yan Zhang.
\newblock Is chatgpt a good recommender? a preliminary study.
\newblock {\em arXiv preprint arXiv:2304.10149}, 2023.

\bibitem{gao2020making}
Tianyu Gao, Adam Fisch, and Danqi Chen.
\newblock Making pre-trained language models better few-shot learners.
\newblock {\em arXiv preprint arXiv:2012.15723}, 2020.

\bibitem{ge2020understanding}
Yingqiang Ge, Shuya Zhao, Honglu Zhou, Changhua Pei, Fei Sun, Wenwu Ou, and Yongfeng Zhang.
\newblock Understanding echo chambers in e-commerce recommender systems.
\newblock In {\em Proceedings of the 43rd international ACM SIGIR conference on research and development in information retrieval}, pages 2261--2270, 2020.

\bibitem{huang2021sliding}
Yanhua Huang, Weikun Wang, Lei Zhang, and Ruiwen Xu.
\newblock Sliding spectrum decomposition for diversified recommendation.
\newblock In {\em Proceedings of the 27th ACM SIGKDD conference on knowledge discovery \& data mining}, pages 3041--3049, 2021.

\bibitem{ziegler2005improving}
Cai-Nicolas Ziegler, Sean~M McNee, Joseph~A Konstan, and Georg Lausen.
\newblock Improving recommendation lists through topic diversification.
\newblock In {\em Proceedings of the 14th international conference on World Wide Web}, pages 22--32, 2005.

\bibitem{ashkan2015optimal}
Azin Ashkan, Branislav Kveton, Shlomo Berkovsky, and Zheng Wen.
\newblock Optimal greedy diversity for recommendation.
\newblock In {\em IJCAI}, volume~15, pages 1742--1748, 2015.

\bibitem{chen2018fast}
Laming Chen, Guoxin Zhang, and Eric Zhou.
\newblock Fast greedy map inference for determinantal point process to improve recommendation diversity.
\newblock {\em Advances in Neural Information Processing Systems}, 31, 2018.

\bibitem{zheng2021dgcn}
Yu~Zheng, Chen Gao, Liang Chen, Depeng Jin, and Yong Li.
\newblock Dgcn: Diversified recommendation with graph convolutional networks.
\newblock In {\em Proceedings of the web conference 2021}, pages 401--412, 2021.

\bibitem{vargas2014coverage}
Sa{\'u}l Vargas, Linas Baltrunas, Alexandros Karatzoglou, and Pablo Castells.
\newblock Coverage, redundancy and size-awareness in genre diversity for recommender systems.
\newblock In {\em Proceedings of the 8th ACM Conference on Recommender systems}, pages 209--216, 2014.

\bibitem{vargas2011intent}
Saul Vargas, Pablo Castells, and David Vallet.
\newblock Intent-oriented diversity in recommender systems.
\newblock In {\em Proceedings of the 34th international ACM SIGIR conference on Research and development in Information Retrieval}, pages 1211--1212, 2011.

\bibitem{shi2012adaptive}
Yue Shi, Xiaoxue Zhao, Jun Wang, Martha Larson, and Alan Hanjalic.
\newblock Adaptive diversification of recommendation results via latent factor portfolio.
\newblock In {\em Proceedings of the 35th international ACM SIGIR conference on Research and development in information retrieval}, pages 175--184, 2012.

\bibitem{wu2019pd}
Qiong Wu, Yong Liu, Chunyan Miao, Binqiang Zhao, Yin Zhao, and Lu~Guan.
\newblock Pd-gan: Adversarial learning for personalized diversity-promoting recommendation.
\newblock In {\em IJCAI}, volume~19, pages 3870--3876, 2019.

\bibitem{10823618}
Wei Xu, Jue Xiao, and Jianlong Chen.
\newblock Leveraging large language models to enhance personalized recommendations in e-commerce.
\newblock In {\em 2024 International Conference on Electrical, Communication and Computer Engineering (ICECCE)}, pages 1--6, 2024.

\bibitem{chen2024dlcrec}
Jiaju Chen, Chongming Gao, Shuai Yuan, Shuchang Liu, Qingpeng Cai, and Peng Jiang.
\newblock Dlcrec: A novel approach for managing diversity in llm-based recommender systems.
\newblock {\em arXiv preprint arXiv:2408.12470}, 2024.

\bibitem{han2024controlling}
Gwangseok Han, Wonbin Kweon, Minsoo Kim, and Hwanjo Yu.
\newblock Controlling diversity at inference: Guiding diffusion recommender models with targeted category preferences.
\newblock {\em arXiv preprint arXiv:2411.11240}, 2024.

\bibitem{carraro2024enhancing}
Diego Carraro and Derek Bridge.
\newblock Enhancing recommendation diversity by re-ranking with large language models.
\newblock {\em ACM Transactions on Recommender Systems}, 2024.

\bibitem{izawa2025diversifying}
Shin Izawa, Keiko Ono, and Panagiotis Adamidis.
\newblock Diversifying furniture recommendations: A user-profile-enhanced recommender vae approach.
\newblock {\em Applied Sciences}, 15(5):2761, 2025.

\bibitem{clarke2008novelty}
Charles~LA Clarke, Maheedhar Kolla, Gordon~V Cormack, Olga Vechtomova, Azin Ashkan, Stefan B{\"u}ttcher, and Ian MacKinnon.
\newblock Novelty and diversity in information retrieval evaluation.
\newblock In {\em Proceedings of the 31st annual international ACM SIGIR conference on Research and development in information retrieval}, pages 659--666, 2008.

\bibitem{zhang2008avoiding}
Mi~Zhang and Neil Hurley.
\newblock Avoiding monotony: improving the diversity of recommendation lists.
\newblock In {\em Proceedings of the 2008 ACM conference on Recommender systems}, pages 123--130, 2008.

\bibitem{bellogin2010study}
Alejandro Bellog{\'\i}n, Iv{\'a}n Cantador, and Pablo Castells.
\newblock A study of heterogeneity in recommendations for a social music service.
\newblock In {\em Proceedings of the 1st International Workshop on Information Heterogeneity and Fusion in Recommender Systems}, pages 1--8, 2010.

\bibitem{Vargas2015}
Saúl Vargas.
\newblock {\em Novelty and Diversity Evaluation and Enhancement in Recommender Systems}.
\newblock Phd thesis, Universidad Autónoma de Madrid, Spain, 2015.
\newblock PhD Dissertation.

\bibitem{puthiya2016coverage}
Shameem~A Puthiya~Parambath, Nicolas Usunier, and Yves Grandvalet.
\newblock A coverage-based approach to recommendation diversity on similarity graph.
\newblock In {\em Proceedings of the 10th ACM Conference on Recommender Systems}, pages 15--22, 2016.

\bibitem{tokutake2023serendipity}
Yu~Tokutake and Kazushi Okamoto.
\newblock Serendipity-oriented recommender system with dynamic unexpectedness prediction.
\newblock In {\em 2023 IEEE International Conference on Systems, Man, and Cybernetics (SMC)}, pages 1247--1252. IEEE, 2023.

\bibitem{park2022psr}
Hyeseong Park, Kyung~Whan Oh, and Uran Oh.
\newblock Psr: Probabilistic serendipitous recommendations.
\newblock In {\em 2022 International Conference on Computational Science and Computational Intelligence (CSCI)}, pages 790--795. IEEE, 2022.

\bibitem{xu2023gs}
Yuanbo Xu, En~Wang, Yongjian Yang, and Hui Xiong.
\newblock Gs\textsuperscript{2}-rs: A generative approach for alleviating cold start and filter bubbles in recommender systems.
\newblock {\em IEEE Transactions on Knowledge and Data Engineering}, 2023.

\bibitem{hassan2024gczrec}
Syed Zain~Ul Hassan, Muhammad Rafi, and Jaroslav Frnda.
\newblock Gczrec: Generative collaborative zero-shot framework for cold start news recommendation.
\newblock {\em IEEE Access}, 2024.

\bibitem{fu2024art}
Zhe Fu and Xi~Niu.
\newblock The art of asking: Prompting large language models for serendipity recommendations.
\newblock In {\em Proceedings of the 2024 ACM SIGIR International Conference on Theory of Information Retrieval}, pages 157--166, 2024.

\bibitem{xi2025bursting}
Yunjia Xi, Muyan Weng, Wen Chen, Chao Yi, Dian Chen, Gaoyang Guo, Mao Zhang, Jian Wu, Yuning Jiang, Qingwen Liu, et~al.
\newblock Bursting filter bubble: Enhancing serendipity recommendations with aligned large language models.
\newblock {\em arXiv preprint arXiv:2502.13539}, 2025.

\bibitem{tokutake2024can}
Yu~Tokutake and Kazushi Okamoto.
\newblock Can large language models assess serendipity in recommender systems?
\newblock {\em Journal of Advanced Computational Intelligence and Intelligent Informatics}, 28(6):1263--1272, 2024.

\bibitem{guan2024review}
Feng Guan and Daisuke Kitayama.
\newblock Review prediction using large-scale language models for serendipity-oriented tourist spot recommendation and its evaluation.
\newblock In {\em 2024 18th International Conference on Ubiquitous Information Management and Communication (IMCOM)}, pages 1--4. IEEE, 2024.

\bibitem{sah2024unveiling}
Chandan~Kumar Sah, Lian Xiaoli, and Muhammad~Mirajul Islam.
\newblock Unveiling bias in fairness evaluations of large language models: A critical literature review of music and movie recommendation systems.
\newblock {\em arXiv preprint arXiv:2401.04057}, 2024.

\bibitem{hua2023tutorial}
Wenyue Hua, Lei Li, Shuyuan Xu, Li~Chen, and Yongfeng Zhang.
\newblock Tutorial on large language models for recommendation.
\newblock In {\em Proceedings of the 17th ACM Conference on Recommender Systems}, pages 1281--1283, 2023.

\bibitem{wei2025addressing}
Xiahua Wei, Naveen Kumar, and Han Zhang.
\newblock Addressing bias in generative ai: Challenges and research opportunities in information management.
\newblock {\em arXiv preprint arXiv:2502.10407}, 2025.

\bibitem{zhang2024gan}
Kangzhi Zhang, Xuezhong Qian, and Wei Song.
\newblock Gan-based fairness-aware recommendation for enhancing the fairness of datagan-based fairness-aware recommendationfairness-aware recommendation.
\newblock In {\em Proceedings of the 2024 Guangdong-Hong Kong-Macao Greater Bay Area International Conference on Digital Economy and Artificial Intelligence}, pages 315--321, 2024.

\bibitem{sakib2024challenging}
Shahnewaz~Karim Sakib and Anindya~Bijoy Das.
\newblock Challenging fairness: A comprehensive exploration of bias in llm-based recommendations.
\newblock In {\em 2024 IEEE International Conference on Big Data (BigData)}, pages 1585--1592. IEEE, 2024.

\bibitem{deldjoo2024understanding}
Yashar Deldjoo.
\newblock Understanding biases in chatgpt-based recommender systems: Provider fairness, temporal stability, and recency.
\newblock {\em ACM Transactions on Recommender Systems}, 2024.

\bibitem{borges2019enhancing}
Rodrigo Borges and Kostas Stefanidis.
\newblock Enhancing long term fairness in recommendations with variational autoencoders.
\newblock In {\em Proceedings of the 11th international conference on management of digital ecosystems}, pages 95--102, 2019.

\bibitem{lilienthal2024synthetic}
Derek~B Lilienthal.
\newblock Synthetic data generation for accurate, fair, and private recommender systems.
\newblock Master's thesis, San Jose State University, 2024.

\bibitem{zhang2023chatgpt}
Jizhi Zhang, Keqin Bao, Yang Zhang, Wenjie Wang, Fuli Feng, and Xiangnan He.
\newblock Is chatgpt fair for recommendation? evaluating fairness in large language model recommendation.
\newblock In {\em Proceedings of the 17th ACM Conference on Recommender Systems}, pages 993--999, 2023.

\bibitem{deldjoo2024cfairllm}
Yashar Deldjoo and Tommaso Di~Noia.
\newblock Cfairllm: Consumer fairness evaluation in large-language model recommender system.
\newblock {\em arXiv preprint arXiv:2403.05668}, 2024.

\bibitem{zhu2021fairness}
Ziwei Zhu, Jianling Wang, and James Caverlee.
\newblock Fairness-aware personalized ranking recommendation via adversarial learning.
\newblock {\em arXiv preprint arXiv:2103.07849}, 2021.

\bibitem{vcutura2024fairness}
Lucija {\v{C}}utura, Klemo Vladimir, Goran Dela{\v{c}}, and Marin {\v{S}}ili{\'c}.
\newblock Fairness in graph-based recommendation: Methods overview.
\newblock In {\em 2024 47th MIPRO ICT and Electronics Convention (MIPRO)}, pages 850--855. IEEE, 2024.

\bibitem{rahman2019fairwalk}
Tahleen Rahman, Bartlomiej Surma, Michael Backes, and Yang Zhang.
\newblock Fairwalk: Towards fair graph embedding.
\newblock 2019.

\bibitem{rao2023can}
Haocong Rao, Cyril Leung, and Chunyan Miao.
\newblock Can chatgpt assess human personalities? a general evaluation framework.
\newblock {\em arXiv preprint arXiv:2303.01248}, 2023.

\bibitem{kliachkin2024fairness1}
Andrii Kliachkin, Eleni Psaroudaki, Jakub Mare{\v{c}}ek, and Dimitris Fotakis.
\newblock Fairness in ranking: Robustness through randomization without the protected attribute.
\newblock In {\em 2024 IEEE 40th International Conference on Data Engineering Workshops (ICDEW)}, pages 201--208. IEEE, 2024.

\bibitem{gao2024brief}
Zhengjie Gao, Xuanzi Liu, Yuanshuai Lan, and Zheng Yang.
\newblock A brief survey on safety of large language models.
\newblock {\em Journal of computing and information technology}, 32(1):47--64, 2024.

\bibitem{lilienthal2024multi}
Derek Lilienthal, Paul Mello, Magdalini Eirinaki, and Stas Tiomkin.
\newblock Multi-resolution diffusion for privacy-sensitive recommender systems.
\newblock {\em IEEE Access}, 2024.

\bibitem{zhang2024lorec}
Kaike Zhang, Qi~Cao, Yunfan Wu, Fei Sun, Huawei Shen, and Xueqi Cheng.
\newblock Lorec: Combating poisons with large language model for robust sequential recommendation.
\newblock In {\em Proceedings of the 47th International ACM SIGIR Conference on Research and Development in Information Retrieval}, pages 1733--1742, 2024.

\bibitem{saputra2024secure}
Adi Saputra, Erma Suryani, and Nur~Aini Rakhmawati.
\newblock Secure and scalable llm-based recommendation systems: An mlops and security by design.
\newblock In {\em 2024 IEEE International Symposium on Consumer Technology (ISCT)}, pages 623--629. IEEE, 2024.

\bibitem{li2024ganprompt}
Xinyu Li, Chuang Zhao, Hongke Zhao, Likang Wu, and Ming He.
\newblock Ganprompt: Enhancing robustness in llm-based recommendations with gan-enhanced diversity prompts.
\newblock {\em arXiv preprint arXiv:2408.09671}, 2024.

\bibitem{zhang2024stealthy}
Jinghao Zhang, Yuting Liu, Qiang Liu, Shu Wu, Guibing Guo, and Liang Wang.
\newblock Stealthy attack on large language model based recommendation.
\newblock {\em arXiv preprint arXiv:2402.14836}, 2024.

\bibitem{ning2024cheatagent}
Liang-bo Ning, Shijie Wang, Wenqi Fan, Qing Li, Xin Xu, Hao Chen, and Feiran Huang.
\newblock Cheatagent: Attacking llm-empowered recommender systems via llm agent.
\newblock In {\em Proceedings of the 30th ACM SIGKDD Conference on Knowledge Discovery and Data Mining}, pages 2284--2295, 2024.

\bibitem{liu2025novel}
Shuangyu Liu, Siyang Yu, Huan Li, Zhibang Yang, Mingxing Duan, and Xiangke Liao.
\newblock A novel shilling attack on black-box recommendation systems for multiple targets.
\newblock {\em Neural Computing and Applications}, 37(5):3399--3417, 2025.

\bibitem{zhiheng2023safety}
Xi~Zhiheng, Zheng Rui, and Gui Tao.
\newblock Safety and ethical concerns of large language models.
\newblock In {\em Proceedings of the 22nd Chinese National Conference on Computational Linguistics (Volume 4: Tutorial Abstracts)}, pages 9--16, 2023.

\bibitem{yao2024survey}
Yifan Yao, Jinhao Duan, Kaidi Xu, Yuanfang Cai, Zhibo Sun, and Yue Zhang.
\newblock A survey on large language model (llm) security and privacy: The good, the bad, and the ugly.
\newblock {\em High-Confidence Computing}, page 100211, 2024.

\bibitem{chua2024ai}
Jaymari Chua, Yun Li, Shiyi Yang, Chen Wang, and Lina Yao.
\newblock Ai safety in generative ai large language models: A survey.
\newblock {\em arXiv preprint arXiv:2407.18369}, 2024.

\bibitem{gao2024novel}
Lina Gao, Jiguo Yu, Jianli Zhao, and Chunqiang Hu.
\newblock A novel temporal privacy-preserving model for social recommendation.
\newblock {\em IEEE Transactions on Computational Social Systems}, 2024.

\bibitem{zhang2024misinformation}
Mengyi Zhang, Qingxing Dong, and Xiaozhen Wu.
\newblock How misinformation diffuses on online social networks: Radical opinions, adaptive relationship, and algorithmic intervention.
\newblock {\em IEEE Transactions on Computational Social Systems}, 2024.

\bibitem{sanner2023large}
Scott Sanner, Krisztian Balog, Filip Radlinski, Ben Wedin, and Lucas Dixon.
\newblock Large language models are competitive near cold-start recommenders for language-and item-based preferences.
\newblock In {\em Proceedings of the 17th ACM conference on recommender systems}, pages 890--896, 2023.

\bibitem{wu2024could}
Xuansheng Wu, Huachi Zhou, Yucheng Shi, Wenlin Yao, Xiao Huang, and Ninghao Liu.
\newblock Could small language models serve as recommenders? towards data-centric cold-start recommendation.
\newblock In {\em Proceedings of the ACM Web Conference 2024}, pages 3566--3575, 2024.

\bibitem{shi2023relieving}
Xiaoyu Shi, Quanliang Liu, Hong Xie, Di~Wu, Bo~Peng, MingSheng Shang, and Defu Lian.
\newblock Relieving popularity bias in interactive recommendation: A diversity-novelty-aware reinforcement learning approach.
\newblock {\em ACM Transactions on Information Systems}, 42(2):1--30, 2023.

\bibitem{xie2021hierarchical}
Ruobing Xie, Shaoliang Zhang, Rui Wang, Feng Xia, and Leyu Lin.
\newblock Hierarchical reinforcement learning for integrated recommendation.
\newblock In {\em Proceedings of the AAAI conference on artificial intelligence}, volume~35, pages 4521--4528, 2021.

\bibitem{sharma2024optimizing}
Amit Sharma, Hua Li, Xue Li, and Jian Jiao.
\newblock Optimizing novelty of top-k recommendations using large language models and reinforcement learning.
\newblock In {\em Proceedings of the 30th ACM SIGKDD Conference on Knowledge Discovery and Data Mining}, pages 5669--5679, 2024.

\bibitem{wu2024survey2}
Haotian Wu, Yingpeng Du, Zhu Sun, Tianjun Wei, Jie Zhang, and Ong~Yew Soon.
\newblock A survey on efficient solutions of large language models for recommendation.
\newblock {\em Authorea Preprints}, 2024.

\bibitem{deldjoo2021survey}
Yashar Deldjoo, Tommaso~Di Noia, and Felice~Antonio Merra.
\newblock A survey on adversarial recommender systems: from attack/defense strategies to generative adversarial networks.
\newblock {\em ACM Computing Surveys (CSUR)}, 54(2):1--38, 2021.

\bibitem{gulla2017adressa}
Jon~Atle Gulla, Lemei Zhang, Peng Liu, {\"O}zlem {\"O}zg{\"o}bek, and Xiaomeng Su.
\newblock The adressa dataset for news recommendation.
\newblock In {\em Proceedings of the international conference on web intelligence}, pages 1042--1048, 2017.

\end{thebibliography}
% \printbibliography % 生成参考文献列表
% \bibliographystyle{unsrt} 
% biography section
% 
% If you have an EPS/PDF photo (graphicx package needed) extra braces are
% needed around the contents of the optional argument to biography to prevent
% the LaTeX parser from getting confused when it sees the complicated
% \includegraphics command within an optional argument. (You could create
% your own custom macro containing the \includegraphics command to make things
% simpler here.)
%\begin{IEEEbiography}[{\includegraphics[width=1in,height=1.25in,clip,keepaspectratio]{mshell}}]{Michael Shell}
% or if you just want to reserve a space for a photo:

% \begin{IEEEbiography}{Michael Shell}
% Biography text here.
% \end{IEEEbiography}

% if you will not have a photo at all:

% \begin{IEEEbiographynophoto}{Shanshan Feng}
% is currently a senior scientist with the Centre for Frontier AI Research (CFAR) and Institute of High Performance Computing (IHPC), Agency for Science, Technology and Research (A*STAR), Singapore.  He received his Ph.D. degree from Nanyang Technological University, Singapore, in 2017. His current research interests include recommendation systems,  graph learning, and generative AI techniques.

% \end{IEEEbiographynophoto}

% \bibliographystyle{ieeetr}
% \bibliography{ref}
 
\end{document}